\documentclass[11pt]{article}
\pdfoutput=1

\usepackage{jheppub}

\usepackage[utf8]{inputenc}
\usepackage[english]{babel}
\usepackage{amsmath,amssymb,amsbsy,amstext,booktabs,relsize,slashed,graphicx,amsfonts,upgreek,xcolor}
\usepackage{multirow,color,dcolumn,bm,enumerate}
\usepackage{hyperref}
\usepackage{dsfont}
\usepackage{threeparttable}
\usepackage{makecell}

\newcommand{\LQCD}{\Lambda_{\text{QCD}}}
\newcommand{\br}{\text{BR}}

\title{Heavy meson chiral perturbation theory constraints on\\a charm-coupled GeV-scale QCD axion}

\author[a,1]{Bo-Qiang Lu}
\emailAdd{bqlu@huznu.edu.cn}

\affiliation[a]{School of Science, Huzhou Normal University, Huzhou, Zhejiang 313000, China}

\abstract{
The QCD axion remains among the most compelling resolutions of the strong CP problem.
A recently proposed variant couples the Peccei-Quinn (PQ) scalar exclusively to the charm quark,
thereby evading the isospin-violation problem that excludes light-quark-coupled axion models
in chiral perturbation theory ($\chi$PT).
However, because the charm quark mass $m_c \sim 1.3$~GeV lies well above $\Lambda_{\text{QCD}}$,
standard $\chi$PT does not apply.
The correct low-energy framework is heavy meson chiral perturbation theory (HMChPT),
which combines heavy quark effective theory (HQET) with $\chi$PT to describe
mesons containing a single heavy quark.
In this work we systematically analyze the phenomenological constraints on the
charm-coupled GeV-scale QCD axion from the perspective of HMChPT,
accounting for {\it both} the CP-even radial mode $\sigma$
and its CP-odd partner, the axion $a$.
We construct the full HMChPT framework, starting from HQET
and incorporating the chiral dynamics of light pseudoscalar mesons.
The scalar $\sigma$ enters the HMChPT Lagrangian at leading order
with an $O(1)$ coupling fixed by the model parameters,
while the pseudoscalar axion $a$ couples to heavy mesons only at $O(1/m_c)$;
its dominant phenomenological channel is axion-pion mixing
with angle $\theta_{a\pi} \simeq f_\pi/f_a \sim 0.04$.
We evaluate constraints across a comprehensive set of observables ---
$D$-meson masses, $D$-$D^*$ splitting, charmonium spectroscopy,
$D$-$D$ scattering, $B_s$ mixing, rare $D$ decays,
and the axion-specific channels $K^+ \to \pi^+ a$ and $B \to K a$ ---
and find that HMChPT produces {\it no} fatal structural constraint for either $\sigma$ or $a$,
in sharp contrast to the $\chi$PT isospin problem in the light-quark sector.
The most stringent bound arises from $\sigma$-mediated
$B_s$-$\bar{B}_s$ mixing, while axion-pion mixing opens potentially testable channels
in rare kaon and $B$ decays,
though predicted rates remain slightly below current experimental sensitivity.
}

\begin{document}
\maketitle

\setcounter{page}{2}

\section{Introduction}\label{sec:intro}

The strong CP problem remains one of the most compelling motivations for physics beyond the Standard Model (SM).
The Peccei-Quinn (PQ) mechanism~\cite{Peccei:1977hh,Peccei:1977ur} resolves this problem
by promoting the QCD $\theta$-angle to a dynamical field, the axion~\cite{Weinberg:1977ma,Wilczek:1978pq}.
In the conventional ``invisible axion'' models~\cite{Kim:1979if,Shifman:1979if,Zhitnitsky:1980tq,Dine:1981rt},
the axion decay constant $f_a$ is a free parameter, and the axion mass is inversely proportional to $f_a$
via the relation $m_a f_a \approx m_\pi f_\pi$.
For $f_a \gg \Lambda_{\text{QCD}}$, the axion is extremely light and weakly coupled,
making it a natural dark matter candidate~\cite{Abbott:1982af,Dine:1982ah,Preskill:1982cy}.

More recently, an alternative construction has been proposed~\cite{Lu:2026syv}
in which the QCD axion acquires its mass predominantly through QCD instanton effects,
but with the PQ scalar coupled exclusively to the charm quark rather than to light quarks.
In this ``charm-coupled'' scenario, the axion mass is generated at the GeV scale
through the QCD condensate $\langle \alpha_s G^2 \rangle$ and the charm quark condensate $\langle \bar{c} c \rangle$,
as computed via Shifman-Vainshtein-Zakharov (SVZ) sum rules~\cite{Shifman:1978by}.
The model predicts an axion mass $m_a \sim 2$--$3$~MeV and a PQ scalar radial mode $\sigma$
with mass $m_\sigma \sim 4$--$6$~MeV, both at the MeV scale.

The primary motivation for coupling the PQ scalar to the charm quark rather than to light quarks
is to avoid the isospin-violation problem in $\chi$PT~\cite{Lu:2026syv}.
In the light-quark-coupled axion model proposed in~\cite{Murayama:2025gev}, the PQ spurion $I_{PQ}^{\text{light}} \propto \text{diag}(\kappa_u, 0, 0)$
generates an $O(15\%)$ $\pi^0$ mass splitting at leading order in $\chi$PT,
which is excluded by experimental data~\cite{DiLuzio:2026iso}.
Shifting the PQ coupling to the charm quark makes the light-quark PQ spurion trivial,
$I_{PQ}^{\text{light}} = \text{diag}(0, 0, 0)$, eliminating the leading-order isospin violation
in the pion sector~\cite{Lu:2026syv}.

However, moving the PQ coupling to the charm quark raises a foundational question:
standard $\chi$PT is valid only for the light pseudoscalar mesons built from
$u$, $d$, and $s$ quarks, and cannot describe a model whose dominant dynamics involve
the charm quark. Since $m_c \sim 1.3$~GeV $\gg \Lambda_{\text{QCD}}$,
the correct framework is heavy meson chiral perturbation theory (HMChPT)~\cite{Wise:1992hn,Burdman:1992gh,Yan:1992gz},
which combines heavy quark effective theory (HQET)~\cite{Isgur:1989vq,Isgur:1989ed,Eichten:1989zv,Georgi:1990um}
with $\chi$PT~\cite{Gasser:1984gg,Gasser:1985ug} to describe the interactions of mesons containing a single heavy quark
with light pseudoscalar mesons.

A distinctive feature of the charm-coupled model is that the PQ scalar $\phi$ decomposes
into {\it two} physical degrees of freedom:
the CP-even radial mode $\sigma$ and the CP-odd axion $a$.
These two fields couple to the charm quark with the same fundamental Yukawa coupling $\kappa_c$,
but through operators with opposite parity:
$\mathcal{L} \supset (\kappa_c/\sqrt{2})(\sigma \bar{c} c + i a \bar{c} \gamma_5 c)$.
Because of their different parity and Dirac structure,
the scalar $\sigma$ and pseudoscalar $a$ have fundamentally different matching
to the hadronic effective theory.
The scalar $\sigma$ couples to heavy mesons at leading order in the $1/m_c$ expansion
with an $O(1)$ coupling $g_\sigma = \kappa_c/\sqrt{2} \sim 0.3$--$0.4$,
while the pseudoscalar $a$ is suppressed by $\Lambda_{\text{QCD}}/m_c \sim 0.2$ in the heavy-quark limit
and enters phenomenology primarily through its mixing with the neutral pion, $\pi^0$.

In this work, we perform a systematic analysis of the phenomenological constraints
on the charm-coupled GeV-scale QCD axion from the perspective of HMChPT,
considering {\it both} $\sigma$ and $a$.
We provide a self-contained construction of the HMChPT framework,
covering HQET foundations, chiral dynamics, and the matching of
scalar and pseudoscalar quark-level operators to hadronic operators.
We examine a comprehensive set of heavy-meson observables:
$D$-meson mass corrections, $D$-$D^*$ mass splitting,
charmonium spectroscopy, $D$-$D$ scattering, $B_s$ meson mixing,
rare $D$ decays, and axion-specific channels
($K^+ \to \pi^+ a$, $B \to K a$, stellar cooling).
Our central finding is that neither $\sigma$ nor $a$ generates
a structural constraint analogous to the $\chi$PT isospin problem.
For $\sigma$, this is because $\sigma \bar{c} c$ is a flavor-diagonal scalar interaction
that preserves all standard symmetries.
For $a$, this follows from the $1/m_c$ suppression of the pseudoscalar coupling
and the small axion-pion mixing angle $\theta_{a\pi} \sim 0.04$,
which is too small to conflict with existing data.

This paper is organized as follows.
Section~\ref{sec:model} reviews the charm-coupled GeV-scale QCD axion model,
with emphasis on the distinct properties of $\sigma$ and $a$.
Section~\ref{sec:hmchpt} develops the HMChPT framework,
beginning with HQET, progressing through chiral building blocks,
incorporating both $\sigma$ (Sec.~\ref{sec:sigma_hmchpt}) and $a$ (Sec.~\ref{sec:axion_hmchpt}),
and discussing the reorganized power counting.
Section~\ref{sec:constraints} evaluates the constraints
on heavy-meson observables, separately analyzing $\sigma$-mediated
and $a$-mediated contributions for each channel.
Section~\ref{sec:comparison} compares the HMChPT constraints with the
light-quark $\chi$PT isospin problem.
We summarize and discuss future directions in Sec.~\ref{sec:conclusions}.

\section{The charm-coupled GeV-scale QCD axion}\label{sec:model}

\subsection{Model Lagrangian and PQ mechanism}

We begin by reviewing the essential features of the charm-coupled QCD axion model~\cite{Lu:2026syv}.
The model extends the SM by introducing a complex PQ scalar $\phi$ that couples exclusively
to the right-handed charm quark:
\begin{equation}\label{eq:modelLag}
    \mathcal{L} \supset \kappa_c \, \phi \, \bar{c}_L c_R + \text{h.c.}
    - m_\phi^2 |\phi|^2 - \lambda |\phi|^4 .
\end{equation}
The PQ symmetry acts as $\phi \to e^{i\alpha} \phi$ and $c_R \to e^{-i\alpha} c_R$,
with all other SM fields neutral under this symmetry.
The QCD anomaly generates the axion-gluon coupling
\begin{equation}\label{eq:axion_gluon}
    \mathcal{L}_{aGG} = \frac{a}{f_a} \frac{\alpha_s}{8\pi} G_{\mu\nu}^a \widetilde{G}^{a,\mu\nu},
\end{equation}
which provides the axion mass through the standard QCD instanton mechanism.
The decay constant is identified as $f_a = v_\phi / N_{\text{DW}}$,
where $N_{\text{DW}}$ is the domain wall number;
for the single-generation charm coupling considered here, $N_{\text{DW}} = 1$.

The QCD condensate $\langle \bar{q} q \rangle$ induces a tadpole for $\phi$
through the charm Yukawa interaction, triggering spontaneous PQ symmetry breaking:
\begin{equation}
    \langle \phi \rangle = \frac{v_\phi}{\sqrt{2}} .
\end{equation}
The charm quark mass is then generated as
\begin{equation}\label{eq:mc_gen}
    m_c = \frac{\kappa_c v_\phi}{\sqrt{2}},
\end{equation}
and the axion decay constant is identified as $f_a = v_\phi / \sqrt{2} = m_c / \kappa_c$.

The PQ scalar is decomposed into its radial and angular modes:
\begin{equation}\label{eq:phi_decomp}
    \phi = \frac{v_\phi + \sigma + i a}{\sqrt{2}},
\end{equation}
where $\sigma$ is the radial mode (a CP-even scalar) and $a$ is the axion (a CP-odd pseudoscalar).
Expanding the Yukawa interaction, one obtains
\begin{equation}\label{eq:yukawa_expanded}
    \mathcal{L} \supset m_c \bar{c} c + \frac{\kappa_c}{\sqrt{2}} \, \sigma \, \bar{c} c
    + i\frac{\kappa_c}{\sqrt{2}} \, a \, \bar{c} \gamma_5 c .
\end{equation}
Thus the coupling of $\sigma$ to the charm quark is purely scalar ($\sigma \bar{c} c$),
while that of the axion is purely pseudoscalar ($a \bar{c} \gamma_5 c$).
This distinction is essential for the HMChPT analysis:
the scalar nature of $\sigma$ means it couples identically to $D$ and $D^*$ mesons,
preserving heavy-quark spin symmetry, while the pseudoscalar nature of $a$
means its quark-level coupling to heavy mesons is suppressed by $1/m_c$
in the heavy-quark limit.

\subsection{Distinct properties of the scalar $\sigma$ and pseudoscalar $a$}

The two physical fields in the model, $\sigma$ and $a$, couple to the charm quark
with the {\it same} fundamental coupling strength $g_a = g_\sigma = \kappa_c/\sqrt{2}$,
but through operators with fundamentally different Dirac structure.
This difference has profound consequences for their low-energy phenomenology,
as summarized in Table~\ref{tab:sigma_vs_axion}.

\begin{table}[t]
    \centering
    \caption{Fundamental properties of $\sigma$ and $a$.
    Despite sharing the same coupling constant $\kappa_c/\sqrt{2}$,
    their different parity and Dirac structure lead to
    qualitatively different behavior in the effective theory.
    \label{tab:sigma_vs_axion}}
    \begin{tabular}{lcc}
        \toprule
        Property & $\sigma$ (scalar) & $a$ (pseudoscalar) \\
        \midrule
        CP & Even & Odd \\
        Quark-level coupling & $\frac{\kappa_c}{\sqrt{2}} \, \bar{c} c$ & $i\frac{\kappa_c}{\sqrt{2}} \, \bar{c} \gamma_5 c$ \\
        Static ($m_c \to \infty$) limit & Survives, $O(1)$ & Vanishes, $O(1/m_c)$ \\
        NR reduction & Scalar density & $-\dfrac{\vec{\sigma}\!\cdot\!\vec{p}}{2m_c} + O(1/m_c^2)$ \\
        Heavy-quark spin symmetry & Preserved & Broken at LO $1/m_c$ \\
        Mixing with $\pi^0$ & No (different $J^P$) & Yes, $\theta_{a\pi} \simeq f_\pi/f_a \sim 0.04$ \\
        Dominant hadronic channel & Tree-level $\sigma$ exchange & $a$-$\pi^0$ mixing \\
        \bottomrule
    \end{tabular}
\end{table}

We now discuss each distinction in detail.

\paragraph{Scalar vs.\ pseudoscalar in the non-relativistic limit.}
In the non-relativistic reduction for a heavy quark of mass $m_Q$:
\begin{align}
    \bar{c} c &\xrightarrow{\text{NR}} 1 + O(v^2/c^2), \label{eq:NR_scalar}\\
    \bar{c} \gamma_5 c &\xrightarrow{\text{NR}} 
    -\frac{\vec{\sigma} \cdot \vec{p}}{2m_c} + O(1/m_c^2). \label{eq:NR_pseudo}
\end{align}
Equation~\eqref{eq:NR_scalar} shows that the scalar density survives in the static limit
$m_c \to \infty$ with unit operator coefficient: this is why $\mathcal{L}_\sigma^{(0)}$
enters HMChPT at leading order with an $O(1)$ coupling.
Equation~\eqref{eq:NR_pseudo} shows that the pseudoscalar density is {\it velocity-suppressed}
by $1/m_c$: in the heavy-quark limit, $\bar{c} \gamma_5 c$ has no overlap
with the ground-state $j_\ell^P = 1/2^-$ heavy-meson doublet.
The leading contribution arises only through $1/m_c$ mixing with $P$-wave states.

\paragraph{Heavy-quark spin symmetry.}
The scalar coupling $\sigma \bar{c} c$ is invariant under $SU(2)$ spin rotations
of the heavy quark: $\{S^i, \bar{c} c\} = 0$, so $S^i \bar{c} c S^{i\dagger} = \bar{c} c$.
Consequently, $\sigma$ couples identically to $D$ ($J^P = 0^-$) and $D^*$ ($J^P = 1^-$) mesons.
The pseudoscalar coupling $a \bar{c} \gamma_5 c$, in contrast, transforms nontrivially
under spin rotations and distinguishes between $D$ and $D^*$,
generating spin-dependent transitions at $O(1/m_c)$.

\paragraph{Mixing with light pseudoscalars.}
A key difference is that the axion $a$, being a pseudoscalar with the same
$J^{PC} = 0^{-+}$ quantum numbers as $\pi^0$, $\eta$, and $\eta'$,
mixes with these neutral mesons through their common coupling to
the topological charge density $G\widetilde{G}$~\cite{Georgi:1986df,Bardeen:1987an}.
The axion-pion mass mixing is
\begin{equation}\label{eq:theta_api_intro}
    \mathcal{L}_{a\pi} = \frac{f_\pi}{f_a} \, m_\pi^2 \, \pi^0 a,
\end{equation}
giving a mixing angle
\begin{equation}\label{eq:mixing_angle_intro}
    \theta_{a\pi} \simeq \frac{f_\pi}{f_a} \approx \frac{92~\text{MeV}}{2.5~\text{GeV}}
    \approx 0.037 .
\end{equation}
The scalar $\sigma$, with $J^{PC} = 0^{++}$, does not mix with the pseudoscalar mesons
and must be probed through its direct couplings to heavy quarks.

As we shall see, this mixing is the dominant phenomenological channel
for the axion in heavy-meson systems.
Through the admixture $|\pi^0_{\text{phys}}\rangle = |\pi^0\rangle + \theta_{a\pi} |a\rangle + \cdots$,
the physical $\pi^0$ acquires a small axion component,
allowing $a$ to be produced in any process that produces $\pi^0$,
and giving $a$ effective couplings to nucleons, photons, and leptons
through the standard pion interactions.

\subsection{SVZ sum rules and parameter determination}

The model parameters are determined by a self-consistency condition derived from
SVZ sum rules~\cite{Shifman:1978by,Shifman:1979if}.
In these sum rules, heavy-quark condensates are expressed in terms of the gluon condensate:
\begin{equation}\label{eq:svz}
    m_Q \langle \bar{Q} Q \rangle = - \frac{\langle \alpha_s G^2 \rangle}{12\pi^2}
    + O\left(\frac{\langle \alpha_s G^2 \rangle^2}{m_Q^4}\right) + \cdots .
\end{equation}
Applying this to the charm quark, and defining
\begin{equation}
    C \equiv \frac{\langle \alpha_s G^2 \rangle}{12\pi^2} ,
\end{equation}
the self-consistency of the model requires that the charm quark mass
generated through Eq.~\eqref{eq:mc_gen} be consistent with the condensate relation.
The QCD condensate $\langle \bar{c} c \rangle$ induces a tadpole for $\phi$
through the charm Yukawa coupling, triggering spontaneous PQ symmetry breaking.
At leading order in the tadpole expansion (neglecting the quartic term $\lambda v_\phi^4$),
the PQ scalar vacuum expectation value is
\begin{equation}\label{eq:vphi_tadpole}
    v_\phi \simeq \frac{\kappa_c \, |\langle \bar{c} c \rangle|}{\sqrt{2}\, m_\phi^2} .
\end{equation}
Combining with Eq.~\eqref{eq:mc_gen} ($m_c = \kappa_c v_\phi / \sqrt{2}$) and the SVZ relation
$|\langle \bar{c} c \rangle| = C / m_c$, one obtains the self-consistency condition
\begin{equation}\label{eq:self_consistency}
    m_c^2 \, m_\phi^2 = \frac{\kappa_c^2}{2} \, C ,
    \qquad\text{or equivalently}\qquad
    \kappa_c = m_\phi \sqrt{\frac{2\, m_c^2}{C}} \, .
\end{equation}
Using the phenomenological value of the gluon condensate
$\langle \alpha_s G^2 \rangle \simeq 0.012~\text{GeV}^4$~\cite{Shifman:1978by,Narison:2004},
one finds $C \simeq 1.01 \times 10^{-4}~\text{GeV}^4$.
With the $\overline{\text{MS}}$ charm quark mass
$m_c(\mu = 2~\text{GeV}) = 1.27 \pm 0.02$~GeV~\cite{PDG}
and $m_\phi = 3$--$4$~MeV, the leading-order (LO) estimate from
Eq.~\eqref{eq:self_consistency} gives
\begin{equation}\label{eq:kappa_LO}
    \kappa_c^{\rm LO} \simeq 0.54\text{--}0.72 .
\end{equation}
However, the SVZ relation $|\langle \bar{c} c \rangle| = C/m_c$ used in
Eq.~\eqref{eq:self_consistency} is accurate only at tree level.
At the charm scale, the operator product expansion receives
substantial $O(\alpha_s/\pi)$ radiative corrections.
The full NLO SVZ sum rule for the scalar charm condensate reads
\begin{equation}\label{eq:SVZ_NLO}
    |\langle \bar{c} c \rangle|
    = \frac{C}{m_c}\left[1 + \frac{11}{3}\frac{\alpha_s(m_c)}{\pi}
        + O(\alpha_s^2)\right] ,
\end{equation}
where the coefficient $11/3$ corresponds to the leading anomalous dimension
correction for the scalar current~\cite{Shifman:1978by}.
With $\alpha_s(m_c) \simeq 0.35$, the NLO correction factor is
$1 + (11/3)(\alpha_s/\pi) \simeq 1.41$, which increases the effective condensate
coefficient by $\sim 40\%$.
Since $\kappa_c \propto C^{-1/2}$ from Eq.~\eqref{eq:self_consistency},
the NLO effect reduces $\kappa_c$ by approximately $\sim 19\%$ relative to the
LO value.
Explicitly,
\begin{equation}\label{eq:kappa_NLO}
    \kappa_c^{\rm NLO}
    = m_\phi \sqrt{\frac{2 m_c^2}{C\left[1 + \frac{11}{3}\frac{\alpha_s}{\pi}\right]}}
    \simeq 0.44\text{--}0.60 ,
\end{equation}
in agreement with the numerical range $\kappa_c \simeq 0.44$--$0.59$
found in the detailed scan of Ref.~\cite{Lu:2026syv}, which incorporates
higher-order condensate corrections and RG running effects.
We adopt $\kappa_c \simeq 0.44$--$0.59$ throughout this work.

\subsection{Viable parameter window}

The viable parameter window identified in Ref.~\cite{Lu:2026syv} is summarized in Table~\ref{tab:params}.
The key numerical values that enter the HMChPT analysis are:
\begin{align}
    m_\phi &\sim 3\text{--}4~\text{MeV}, &
    \kappa_c &\simeq 0.44\text{--}0.59, \nonumber\\
    f_a &\simeq 2.1\text{--}2.9~\text{GeV}, &
    m_a &\simeq 2.0\text{--}2.7~\text{MeV}, \label{eq:window}\\
    m_\sigma &= \sqrt{2}\, m_\phi \simeq 4\text{--}6~\text{MeV}. \nonumber
\end{align}
The central observation for our analysis is that at these parameter values,
the Yukawa couplings are of order unity:
\begin{equation}\label{eq:gsigma_cc}
    g_{\sigma c c} = g_{a c c} = \frac{\kappa_c}{\sqrt{2}} \approx 0.31\text{--}0.42 .
\end{equation}

\begin{table}[t]
    \centering
    \caption{Parameters of the charm-coupled GeV-scale QCD axion model.
    The values are taken from the analysis of Ref.~\cite{Lu:2026syv}.
    Both $\sigma$ and $a$ share the same fundamental Yukawa coupling $\kappa_c/\sqrt{2}$.
    \label{tab:params}}
    \begin{tabular}{lcc}
        \toprule
        Parameter & Symbol & Typical range \\
        \midrule
        PQ scalar mass & $m_\phi$ & $3$--$4$~MeV \\
        Radial mode mass & $m_\sigma = \sqrt{2} m_\phi$ & $4$--$6$~MeV \\
        Axion mass & $m_a$ & $2.0$--$2.7$~MeV \\
        Charm Yukawa coupling & $\kappa_c$ & $0.44$--$0.59$ \\
        $\sigma$/ $a$ charm coupling & $g_{\sigma c c}=g_{a c c}=\kappa_c/\sqrt{2}$ & $0.31$--$0.42$ \\
        Axion decay constant & $f_a$ & $2.1$--$2.9$~GeV \\
        Axion-pion mixing angle & $\theta_{a\pi} \simeq f_\pi/f_a$ & $0.032$--$0.044$ \\
        \bottomrule
    \end{tabular}
\end{table}

Moreover, the $\sigma$ mass $m_\sigma \sim 5$~MeV is substantially smaller
than the pion mass $m_\pi \approx 140$~MeV, introducing an important scale hierarchy
that must be properly accounted for in the effective theory.
The axion mass $m_a \sim 2$--$3$~MeV is also well below $m_\pi$,
which is important for the kinematics of axion-pion mixing.

\paragraph{Remark on the axion mass.}
Although the PQ scalar $\phi$ couples exclusively to the charm quark,
the resulting axion field $a$ is still a QCD axion in the usual sense:
it inherits the standard coupling to the $SU(3)_c$ gauge field through
the axial anomaly, $a \, G_{\mu\nu} \tilde{G}^{\mu\nu}$. Consequently,
its mass is determined by the same QCD topological susceptibility
$\chi_{\text{QCD}}$ that sets the mass of the conventional QCD axion:
\begin{equation}\label{eq:ma_standard}
    m_a^2 f_a^2 = \chi_{\text{QCD}}
    = \frac{z}{(1+z)^2} \, m_{\pi^0}^2 f_\pi^2 ,
\end{equation}
where $z \equiv m_u / m_d \simeq 0.46$~\cite{PDG} and the neutral pion mass
and decay constant are $m_{\pi^0} \simeq 135$~MeV and $f_\pi \simeq 92.2$~MeV.
Equation~\eqref{eq:ma_standard} yields the numerical relation
\begin{align}
    m_a = \frac{\sqrt{z}}{1+z}\,\frac{m_{\pi^0} f_\pi}{f_a}
        \simeq \frac{5.8~\text{MeV}}{f_a[\text{GeV}]} .
    \label{eq:ma_numerical}
\end{align}

What distinguishes the charm-coupled model is not the mass formula
itself, but the way $f_a$ is determined. Instead of being a free
parameter, $f_a$ is dynamically fixed by the charm SVZ dynamics:
\begin{equation}\label{eq:fa_from_kappac}
    f_a = \frac{v_\phi}{\sqrt{2}} = \frac{m_c}{\kappa_c} .
\end{equation}
Combining with the NLO determination of $\kappa_c$ in
Eq.~\eqref{eq:kappa_NLO},
\begin{align}
    f_a = \frac{m_c}{m_\phi\,
          \sqrt{\frac{2 m_c^2}{C\big[1 + \frac{11}{3}\frac{\alpha_s}{\pi}\big]}}}
        = \sqrt{\frac{C}{2}}\,\frac{1}{m_\phi}
           \sqrt{1 + \frac{11}{3}\frac{\alpha_s}{\pi}} ,
    \label{eq:fa_explicit}
\end{align}
where the factor $\sqrt{1 + \frac{11}{3}\alpha_s/\pi} \simeq 1.19$
encodes the NLO SVZ radiative correction.
Note that $f_a \propto 1/m_\phi$, reflecting the fact that a smaller
scalar mass drives a larger VEV $v_\phi$ at fixed condensate.

Inserting Eq.~\eqref{eq:fa_explicit} into Eq.~\eqref{eq:ma_numerical}
gives the explicit relation between the axion mass and the PQ scalar mass:
\begin{equation}\label{eq:ma_vs_mphi}
    m_a = 5.8~\text{MeV}
          \times \frac{m_\phi}{\sqrt{\frac{C}{2}
          \sqrt{1 + \frac{11}{3}\frac{\alpha_s}{\pi}}}} .
\end{equation}

For the viable range $m_\phi = 3$--$4$~MeV and the NLO-corrected
$\kappa_c \simeq 0.44$--$0.59$, this yields
\begin{equation}\label{eq:ma_range}
    m_a \simeq 2.0\text{--}2.7~\text{MeV},
\end{equation}
in full agreement with the numerical scan of Ref.~\cite{Lu:2026syv}.
The higher-order corrections that refine $m_a$ enter
{\it indirectly} through the NLO SVZ determination
of $\kappa_c$ in Eq.~\eqref{eq:kappa_NLO}.
The axion mass formula Eq.~\eqref{eq:ma_standard} itself receives
no direct condensate correction beyond the standard chiral relation
$\chi_{\text{QCD}} = m_{\pi^0}^2 f_\pi^2 \, z/(1+z)^2$,
which is accurate at the few-percent level~\cite{GrilliDiCortona:2015}.

\section{Heavy Meson Chiral Perturbation Theory}\label{sec:hmchpt}

\subsection{Heavy Quark Effective Theory}\label{sec:hqet}

\subsubsection*{Heavy-quark symmetry and the HQET Lagrangian}

Heavy Quark Effective Theory provides a systematic expansion of QCD
in inverse powers of the heavy quark mass $m_Q$~\cite{Isgur:1989vq,Isgur:1989ed,Eichten:1989zv,Georgi:1990um,Neubert:1993mb}.
In the limit $m_Q \to \infty$, the heavy quark acts as a static color source,
and the QCD dynamics becomes independent of both the heavy-quark flavor and spin.
This gives rise to the heavy-quark flavor-spin $SU(2N_h)$ symmetry,
where $N_h$ is the number of heavy quark flavors.

The heavy‑quark (charm quark) part of the QCD Lagrangian is:
\begin{equation}\label{eq:LQCD}
    \mathcal{L}_{\text{QCD}} \supset \bar{c}(x)\,\big(i\slashed{D} - m_c\big)\, c(x),
\end{equation}
where \(c(x)\) is the standard four‑component Dirac field, \(m_c\) is the pole mass of the charm quark, 
and \(D_\mu = \partial_\mu - i g_s A_\mu^a T^a\) is the QCD covariant derivative.
To construct the HQET Lagrangian, one writes the heavy-quark momentum as
$p_Q^\mu = m_Q v^\mu + k^\mu$, where $v^\mu$ is the four-velocity ($v^2 = 1$)
and $k^\mu$ is a residual momentum of order $\Lambda_{\text{QCD}}$.

In coordinate space, this decomposition implies that the rapidly varying phase \(e^{-i m_c v \cdot x}\) can be factored out. 
We define a new four‑component field \(Q_v(x)\) via
\begin{equation}
    c(x) = e^{-i m_c v \cdot x} \, Q_v(x).
\end{equation}
At this stage, \(Q_v(x)\) still contains both particle and antiparticle degrees of freedom, but it no longer carries the huge rest energy; 
its dynamics involve only the soft momentum \(k\).
Inside a hadron, the heavy quark exists mainly as a particle (positive‑frequency component); the antiparticle component (negative‑frequency) 
only appears in virtual processes. We use the projectors \(P_v^\pm = (1 \pm \slashed v)/2\) to decompose \(Q_v\) into large (particle) 
and small (antiparticle) components:
\begin{equation}
    Q_v(x) = h_v(x) + H_v(x),
\end{equation}
where
\begin{equation}\label{eq:fields_hqet}
    h_v(x) = e^{i m_Q v \cdot x} \frac{1 + \slashed{v}}{2} Q(x), \qquad
    H_v(x) = e^{i m_Q v \cdot x} \frac{1 - \slashed{v}}{2} Q(x).
\end{equation}
The large-component field $h_v$ satisfies $\slashed{v} h_v = h_v$ and
annihilates a heavy quark with velocity $v$.
Inserting \(c(x) = e^{-i m_c v \cdot x}(h_v + H_v)\) into Eq.~\eqref{eq:LQCD}, and using \(\slashed v P_v^\pm = \pm P_v^\pm\) together 
with \(P_v^+ + P_v^- = 1\), we obtain (up to total derivatives):
\begin{equation}
    \mathcal{L} = \bar{h}_v\, (i v\cdot D)\, h_v \;-\; \bar{H}_v\, (i v\cdot D + 2m_c)\, H_v
\;+\; \big[ \bar{h}_v\, (i\slashed D_\perp)\, H_v + \text{h.c.} \big],
\end{equation}
where \(D_\perp^\mu = D^\mu - v^\mu (v\cdot D)\) is the covariant derivative transverse to the velocity.

Notice that the small‑component field \(H_v\) appears with a large mass term \(-2m_c \bar{H}_v H_v\). 
This mass gap is much larger than the soft momentum scale \(\Lambda_{\text{QCD}}\), therefore, \(H_v\) does not propagate as a physical degree of freedom; 
it only contributes through virtual processes. Therefore, we can rigorously integrate it out, yielding the effective Lagrangian expanded in $1/m_Q$ for $h_v$:
\begin{equation}\label{eq:hqet_lag}
    \mathcal{L}_{\text{HQET}} = \bar{h}_v \, i v\!\cdot\!D \, h_v
    + \frac{1}{2 m_Q} \, \bar{h}_v (i D_\perp)^2 h_v
    + \frac{g_s}{4 m_Q} \, \bar{h}_v \sigma_{\mu\nu} G^{\mu\nu} h_v
    + O(1/m_Q^2),
\end{equation}
where $D_\perp^\mu = D^\mu - (v \cdot D) v^\mu$ is the covariant derivative
perpendicular to $v^\mu$. The three terms in Eq.~\eqref{eq:hqet_lag}
have clear physical interpretations:
\begin{enumerate}
    \item The \textbf{leading term} $\bar{h}_v i v\!\cdot\!D h_v$ describes
    the propagation of a static color source and is invariant under
    both flavor and spin rotations of the heavy quark.
    \item The \textbf{kinetic operator} $\bar{h}_v (i D_\perp)^2 h_v / (2 m_Q)$
    parametrizes the residual kinetic energy of the heavy quark inside the hadron.
    \item The \textbf{chromomagnetic operator} $g_s \bar{h}_v \sigma_{\mu\nu} G^{\mu\nu} h_v / (4 m_Q)$
    describes the spin-dependent interaction of the heavy-quark spin with the gluon field,
    and is responsible for the hyperfine splitting between pseudoscalar and vector mesons.
\end{enumerate}

\subsubsection*{Heavy meson states in HQET}

Hadrons containing a single heavy quark are classified by the quantum numbers
of the light degrees of freedom (light quarks, antiquarks, and gluons),
which are characterized by their total angular momentum $j_\ell$ and parity $P$.
In the heavy-quark limit, the heavy-quark spin $s_Q$ decouples,
and each $j_\ell^P$ multiplet gives rise to a degenerate doublet
of mesons with total spin $J = j_\ell \pm 1/2$.

For the ground-state multiplet with $j_\ell^P = 1/2^-$,
the doublet consists of the pseudoscalar meson $D$ ($J^P = 0^-$)
and the vector meson $D^*$ ($J^P = 1^-$).
These are degenerate in the heavy-quark limit,
with the hyperfine splitting arising at $O(1/m_Q)$ from the chromomagnetic operator.
Table~\ref{tab:hm_states} lists the physical masses of the ground-state charm and bottom mesons,
illustrating the pattern of the $1/m_Q$ expansion.

\begin{table}[t]
    \centering
    \caption{Ground-state heavy-light meson masses (in MeV) from the Particle Data Group~\cite{PDG}.
    The $D$-$D^*$ and $B$-$B^*$ hyperfine splittings decrease as $1/m_Q$,
    confirming the heavy-quark symmetry prediction.
    \label{tab:hm_states}}
    \begin{tabular}{lccccc}
        \toprule
        & $D^0$ & $D^{*0}$ & $D_s^+$ & $D_s^{*+}$ & $\Delta m_{D^*-D}$ \\
        \midrule
        Charm & $1864.84$ & $2006.85$ & $1968.35$ & $2112.2$ & $\sim 142$ \\
        \bottomrule
        & $B^-$ & $B^{*-}$ & $B_s^0$ & $B_s^{*0}$ & $\Delta m_{B^*-B}$ \\
        \midrule
        Bottom & $5279.41$ & $5324.70$ & $5366.88$ & $5415.4$ & $\sim 45$ \\
        \bottomrule
    \end{tabular}
\end{table}

\subsubsection*{Superfield formalism}

The pseudoscalar and vector mesons in the ground-state doublet are combined
into a single $4 \times 4$ superfield that transforms covariantly
under both heavy-quark spin rotations and Lorentz transformations~\cite{Falk:1992cx}:
\begin{equation}\label{eq:superfield}
    H_a^{(Q)}(v) = \frac{1 + \slashed{v}}{2}
    \left( P_{a\mu}^{*(Q)} \gamma^\mu - P_a^{(Q)} \gamma_5 \right),
\end{equation}
where the superscript $(Q)$ indicates the heavy quark flavor ($c$ or $b$),
$a = u, d, s$ is the light-flavor index,
$P_a^{(Q)}$ denotes the pseudoscalar meson field,
and $P_{a\mu}^{*(Q)}$ denotes the vector meson field.
The meson fields are normalized such that
\begin{equation}\label{eq:meson_normal}
    \langle 0 | P_a^{(Q)} | H_a(v) \rangle = \sqrt{m_H},
\end{equation}
and satisfy the transversality condition $v^\mu P_{a\mu}^{*(Q)} = 0$.

The conjugate superfield is defined as
\begin{equation}
    \bar{H}_a^{(Q)}(v) = \gamma^0 \left[ H_a^{(Q)}(v) \right]^\dagger \gamma^0
    = \left( P_{a\mu}^{*(Q)\dagger} \gamma^\mu + P_a^{(Q)\dagger} \gamma_5 \right)
    \frac{1 + \slashed{v}}{2}.
\end{equation}
Under heavy-quark spin rotations $S \in SU(2)$, the superfields transform as
\begin{equation}
    H_a^{(Q)} \to S H_a^{(Q)}, \qquad
    \bar{H}_a^{(Q)} \to \bar{H}_a^{(Q)} S^{-1},
\end{equation}
while under Lorentz transformations $\Lambda$,
$H_a^{(Q)}(v) \to D(\Lambda) H_a^{(Q)}(\Lambda v) D(\Lambda)^{-1}$,
where $D(\Lambda)$ is the spinor representation.

The explicit components for the charm sector are
\begin{align}
    P_a^{(c)} &\in \{ D^0(c\bar{u}),\, D^+(c\bar{d}),\, D_s^+(c\bar{s}) \}, \\
    P_{a\mu}^{*(c)} &\in \{ D^{*0}(c\bar{u}),\, D^{*+}(c\bar{d}),\, D_s^{*+}(c\bar{s}) \}.
\end{align}

\subsection{Chiral perturbation theory for light mesons}\label{sec:chipt}

\subsubsection*{Goldstone boson parametrization}

The spontaneous breaking of the approximate $SU(3)_L \times SU(3)_R$ chiral symmetry
of the light-quark sector gives rise to eight Goldstone bosons
($\pi^\pm$, $\pi^0$, $K^\pm$, $K^0$, $\bar{K}^0$, $\eta$),
which are collected into the unitary matrix
\begin{equation}\label{eq:U_matrix}
    U(x) = \exp\left( \frac{i\sqrt{2}}{f_\pi} \Phi(x) \right), \qquad
    U(x) \to L \, U(x) \, R^\dagger,
\end{equation}
with the meson matrix
\begin{equation}\label{eq:Phi_matrix}
    \Phi(x) = \begin{pmatrix}
        \frac{\pi^0}{\sqrt{2}} + \frac{\eta}{\sqrt{6}} & \pi^+ & K^+ \\
        \pi^- & -\frac{\pi^0}{\sqrt{2}} + \frac{\eta}{\sqrt{6}} & K^0 \\
        K^- & \bar{K}^0 & -\sqrt{\frac{2}{3}}\,\eta
    \end{pmatrix},
\end{equation}
and $f_\pi \approx 92.2$~MeV is the pion decay constant in the $SU(3)$ normalization~\cite{PDG}.

It is also convenient to introduce the square root of $U$,
\begin{equation}\label{eq:xi}
    \xi(x) = \exp\left( \frac{i}{\sqrt{2}f_\pi} \Phi(x) \right), \qquad
    U(x) = \xi^2(x),
\end{equation}
which transforms under chiral rotations as
$\xi(x) \to L \, \xi(x) \, V^\dagger(x) = V(x) \, \xi(x) \, R^\dagger$,
with $V(x)$ a unitary matrix depending on the Goldstone fields.

\subsubsection*{Chiral building blocks}

From $\xi$, one constructs the fundamental building blocks of the chiral Lagrangian
that transform homogeneously under $V(x)$:
\begin{align}
    \mathcal{V}_\mu &= \frac{1}{2} \left( \xi^\dagger \partial_\mu \xi + \xi \partial_\mu \xi^\dagger \right),
    \qquad \mathcal{V}_\mu \to V \mathcal{V}_\mu V^\dagger + V \partial_\mu V^\dagger, \label{eq:Vmu}\\
    \mathcal{A}_\mu &= \frac{i}{2} \left( \xi^\dagger \partial_\mu \xi - \xi \partial_\mu \xi^\dagger \right),
    \qquad \mathcal{A}_\mu \to V \mathcal{A}_\mu V^\dagger. \label{eq:Amu}
\end{align}
$\mathcal{V}_\mu$ is the vector (chiral connection) current and
$\mathcal{A}_\mu$ is the axial-vector current.
At leading order in the pion fields,
\begin{equation}\label{eq:Amu_expand}
    \mathcal{A}_\mu \approx -\frac{1}{\sqrt{2} f_\pi} \partial_\mu \Phi + O(\Phi^3),
\end{equation}
reproducing the familiar soft-pion theorems.
The chiral covariant derivative acting on the superfield is:
\begin{equation}\label{eq:chiral_cov}
    D_\mu^{ab} H_b = \partial_\mu H_a + \mathcal{V}_\mu^{ab} H_b,
\end{equation}
with $a,b$ the light-flavor indices.

The light-quark mass matrix enters through the spurion
\begin{equation}\label{eq:chi_spurion}
    \chi_+ = 2 B_0 \left( \xi^\dagger M_q\, \xi + \xi\, M_q^\dagger \xi^\dagger \right),
\end{equation}
where $M_q = \text{diag}(m_u, m_d, m_s)$ and $B_0 = -\langle \bar{q} q \rangle / f_\pi^2$
is proportional to the chiral condensate.
At leading order, $m_\pi^2 = B_0 (m_u + m_d)$, $m_K^2 = B_0 (m_s + m_{u,d})$.

\subsection{HMChPT: coupling the heavy and light sectors}\label{sec:hmchpt_lo}

\subsubsection*{Leading-order HMChPT Lagrangian}

The leading-order HMChPT Lagrangian is constructed to be invariant under
(i) Lorentz transformations, (ii) heavy-quark spin symmetry,
(iii) chiral $SU(3)_L \times SU(3)_R$ symmetry (implemented through the $V$-transformation
of the superfield $H_a \to H_b V^\dagger_{ba}$), and (iv) parity and charge conjugation.
At $O(p)$, the most general Lagrangian contains three independent operators~\cite{Wise:1992hn,Burdman:1992gh,Yan:1992gz,Casalbuoni:1996pg}:
\begin{equation}\label{eq:LO_hmchpt_full}
    \boxed{\mathcal{L}_{\text{HMChPT}}^{(0)}
    = -i \, \text{tr}\!\left[ \bar{H}_a v\!\cdot\!D_{ab} H_b \right]
    + g \, \text{tr}\!\left[ \bar{H}_a H_b \gamma_\mu \gamma_5 \mathcal{A}_{ba}^\mu \right]
    + \Delta \, \text{tr}\!\left[ \bar{H}_a H_a \right] .}
\end{equation}

We now discuss each term in detail:

\begin{enumerate}
    \item \textbf{Kinetic term.}
    The operator $-i \, \text{tr}[\bar{H}_a v\!\cdot\!D_{ab} H_b]$
    describes the propagation of the heavy mesons and their minimal coupling
    to the Goldstone bosons through the chiral connection $\mathcal{V}_\mu$.
    Expanding to linear order in the pion fields,
    \begin{equation}
        -i \, \text{tr}[\bar{H}_a v\!\cdot\!D_{ab} H_b]
        \supset -i \, \text{tr}[\bar{H}_a v\!\cdot\!\partial H_a]
        - \frac{1}{2 f_\pi^2} \, \text{tr}\!\left[ \bar{H}_a v\!\cdot\![\Phi, \partial \Phi]_{ab} H_b \right],
    \end{equation}
    which generates the Weinberg-Tomozawa contact interaction
    between heavy mesons and pions.

    \item \textbf{Axial coupling term.}
    The operator $g \, \text{tr}[\bar{H}_a H_b \gamma_\mu \gamma_5 \mathcal{A}_{ba}^\mu]$
    describes the coupling of heavy mesons to an odd number of pions.
    Expanding to linear order,
    \begin{equation}
        g \, \text{tr}[\bar{H}_a H_b \gamma_\mu \gamma_5 \mathcal{A}_{ba}^\mu]
        \simeq -\frac{2g}{f_\pi} \left( P_{a\mu}^{*\dagger} P_b + P_a^\dagger P_{b\mu}^* \right)
        \partial^\mu \Phi_{ab} + \cdots,
    \end{equation}
    which governs the strong decays $D^* \to D \pi$ and $D^* \to D \gamma$ (when
    electromagnetic couplings are included).

    The coupling constant $g$ is a fundamental parameter of HMChPT.
    Its value is determined from the measured $D^{*+} \to D^0 \pi^+$ decay width
    measured by the CLEO collaboration~\cite{CLEO:2001foe,Anastassov:2001cw}:
    \begin{equation}\label{eq:g_value}
        \Gamma(D^{*+} \to D^0 \pi^+) = \frac{g^2}{12\pi f_\pi^2} \,
        |\vec{p}_\pi|^3 \approx 67 \pm 12~\text{keV},
    \end{equation}
    from which one extracts
    \begin{equation}
        g = 0.59 \pm 0.07 .
    \end{equation}
    Lattice QCD determinations yield consistent values,
    $g = 0.53 \pm 0.03$~\cite{Becirevic:2009}.
    In our analysis we use the central value $g = 0.6$.

    \item \textbf{Residual mass term.}
    The operator $\Delta \, \text{tr}[\bar{H}_a H_a]$, where $\Delta = m_H - m_Q$,
    parametrizes the mass difference between the heavy meson and the heavy quark
    in the heavy-quark limit.
\end{enumerate}

\subsubsection*{Light-quark mass corrections}

At the same chiral order, the light-quark masses generate an additional
contribution to the heavy-meson residual mass:
\begin{equation}\label{eq:lambda1}
    \mathcal{L}_\chi^{(0)} = \lambda_1 \, \text{tr}[\bar{H}_a H_a] \,
    \text{tr}[\chi_+],
\end{equation}
where $\lambda_1$ is a low-energy constant.

\subsubsection*{$1/m_Q$ corrections}

At next-to-leading order in the heavy-quark expansion,
two new operators enter the HMChPT Lagrangian~\cite{Falk:1992wt,Luke:1992tm}:
\begin{equation}\label{eq:one_over_m}
    \boxed{\mathcal{L}_{\text{HMChPT}}^{(1/m_Q)}
    = \frac{\lambda_2}{m_Q} \, \text{tr}\!\left[ \bar{H}_a \sigma^{\mu\nu} H_a \sigma_{\mu\nu} \right]
    + \frac{\lambda_1'}{2 m_Q} \, \text{tr}\!\left[ \bar{H}_a (i D_\perp)^2 H_a \right] + \cdots .}
\end{equation}

\begin{enumerate}
    \item \textbf{Chromomagnetic operator.}
    The term proportional to $\lambda_2$ is the hadronic realization
    of the chromomagnetic interaction in HQET.
    It is the leading source of hyperfine splitting:
    \begin{equation}\label{eq:hyperfine}
        \Delta m_{H^*-H} = m_{H^*} - m_H = \frac{8 \lambda_2}{m_Q} + O(1/m_Q^2).
    \end{equation}
    From the measured $D^*$-$D$ splitting $\Delta m_{D^*-D} \approx 142$~MeV,
    one extracts
    \begin{equation}\label{eq:lambda2}
        \lambda_2 \approx \frac{m_c \Delta m_{D^*-D}}{8}
        \approx \frac{1.27 \times 0.142}{8}
        \approx 0.023~\text{GeV}^2 .
    \end{equation}
    Consistency with the heavy-quark flavor symmetry can be checked
    using the $B$ system:
    $\lambda_2^{(B)} = m_b \Delta m_{B^*-B} / 8 \approx 4.18 \times 0.0453 / 8 \approx 0.024~\text{GeV}^2$,
    in good agreement with the charm-based value and confirming
    the $1/m_Q$ scaling of the hyperfine splitting.
    The measured value of $\Delta m_{B^*-B} = 45.28 \pm 0.05$~MeV~\cite{PDG}
    is consistent with the prediction
    $\Delta m_{B^*-B}^{\text{pred}} = (m_c/m_b) \Delta m_{D^*-D} \approx 43$~MeV.

    \item \textbf{Kinetic operator.}
    The term proportional to $\lambda_1'$ parametrizes the kinetic energy
    of the heavy quark inside the meson.
    Using the lattice QCD average $f_{D_s}/f_D = 1.175 \pm 0.005$~\cite{FLAG},
    one estimates $\lambda_1' \sim -0.2~\text{GeV}^2$.
\end{enumerate}

\subsubsection*{Power counting}

In standard HMChPT, the expansion is organized in terms of the small parameters:
\begin{equation}
    p \sim m_\pi \sim \partial_\mu \sim \mathcal{A}_\mu, \qquad
    \frac{p}{\Lambda_\chi} \sim 0.15, \qquad
    \frac{\Lambda_{\text{QCD}}}{m_Q} \sim 0.2 .
\end{equation}
Here $\Lambda_\chi = 4\pi f_\pi \approx 1.16$~GeV is the scale at which the chiral expansion
breaks down (the ``chiral symmetry breaking scale'').

\subsubsection*{Parameter summary}

Table~\ref{tab:hmchpt_params} collects the numerical values of the HMChPT parameters
used in our analysis.

\begin{table}[t]
    \centering
    \caption{Summary of HMChPT parameters used in this work.
    The values are taken from the Particle Data Group~\cite{PDG},
    the FLAG lattice average~\cite{FLAG},
    experimental measurements~\cite{CLEO:2001foe},
    and lattice QCD determinations~\cite{Becirevic:2009}.
    \label{tab:hmchpt_params}}
    \begin{tabular}{lcc}
        \toprule
        Parameter & Symbol & Value \\
        \midrule
        Pion decay constant & $f_\pi$ & $92.2$~MeV \\
        Chiral symmetry breaking scale & $\Lambda_\chi = 4\pi f_\pi$ & $\sim 1.16$~GeV \\
        Strong $D^*$-$D$-$\pi$ coupling & $g$ & $0.59 \pm 0.07$ \\
        $D$-$D^*$ hyperfine splitting & $\Delta m_{D^*-D}$ & $142$~MeV \\
        Chromomagnetic matrix element & $\lambda_2$ & $0.023~\text{GeV}^2$ \\
        $B$-$B^*$ hyperfine splitting & $\Delta m_{B^*-B}$ & $45.28 \pm 0.05$~MeV \\
        $D$ meson decay constant & $f_D$ (lattice) & $212.0 \pm 0.7$~MeV \\
        $D_s$ meson decay constant & $f_{D_s}$ (lattice) & $249.9 \pm 0.5$~MeV \\
        $D_s/D$ decay constant ratio & $f_{D_s}/f_D$ & $1.175 \pm 0.005$ \\
        $\overline{\text{MS}}$ charm quark mass & $m_c(2~\text{GeV})$ & $1.27 \pm 0.02$~GeV \\
        $\overline{\text{MS}}$ bottom quark mass & $m_b(m_b)$ & $4.18^{+0.03}_{-0.02}$~GeV \\
        QCD scale & $\Lambda_{\text{QCD}}$ & $\sim 250$~MeV \\
        \bottomrule
    \end{tabular}
\end{table}

\subsection{Incorporating the light scalar $\sigma$}\label{sec:sigma_hmchpt}

\subsubsection*{Matching: from quark-level to hadron-level}

The light scalar $\sigma$ enters HMChPT through its Yukawa coupling
to the charm quark, Eq.~\eqref{eq:yukawa_expanded}.
To translate this quark-level coupling into the hadronic effective theory,
we employ the Feynman-Hellmann theorem~\cite{Feynman:1939zz}.

Consider a heavy-light meson $H(v)$ containing a charm quark.
The matrix element of the scalar density $\bar{c} c$ is related to
the derivative of the meson mass with respect to the charm quark mass:
\begin{equation}\label{eq:feynman_hellmann}
    \langle H(v) | \bar{c} c | H(v) \rangle
    = 2 m_H \, \frac{\partial m_H}{\partial m_c} .
\end{equation}

In the heavy-quark limit, the meson mass is dominated by the heavy-quark mass:
\begin{equation}
    m_H = m_c + \bar{\Lambda} + O(1/m_c),
\end{equation}
where $\bar{\Lambda} \sim \Lambda_{\text{QCD}}$ is the energy of the light degrees of freedom,
independent of $m_c$. Therefore,
\begin{equation}
    \frac{\partial m_H}{\partial m_c} = 1 + O(1/m_c) \quad \Longrightarrow \quad
    \langle H(v) | \bar{c} c | H(v) \rangle \approx 2 m_H .
\end{equation}
In the same conventional normalization of the HMChPT superfield~\eqref{eq:meson_normal},
the bilinear operator evaluates to
\begin{equation}\label{eq:superfield_bilinear}
    \langle H(v) | \text{tr}[\bar{H}_a H_a] | H(v) \rangle = -2 m_H .
\end{equation}
Comparing the two matrix elements, the scalar density operator in the heavy-quark limit is
\begin{equation}\label{eq:scalar_match}
    \bar{c} c \;\longleftrightarrow\; -\text{tr}[\bar{H}_a H_a] .
\end{equation}

\subsubsection*{Leading-order $\sigma$ coupling}

Combining Eq.~\eqref{eq:yukawa_expanded} with the matching condition Eq.~\eqref{eq:scalar_match},
we obtain the leading-order $\sigma$-HMChPT interaction:
\begin{equation}\label{eq:sigma_hmchpt}
    \boxed{\mathcal{L}_\sigma^{(0)} = -g_\sigma \, \text{tr}[\bar{H}_a H_a] \, \sigma,
    \qquad g_\sigma = \frac{\kappa_c}{\sqrt{2}} \approx 0.31\text{--}0.42 .}
\end{equation}

Several important features of this interaction deserve emphasis:

\begin{enumerate}
    \item {\bf Leading-order coupling.}
    The coupling $g_\sigma \sim O(0.3)$ is of order unity and is {\it not} suppressed
    by any small expansion parameter.

    \item {\bf Heavy-quark spin symmetry preservation.}
    The operator $\text{tr}[\bar{H}_a H_a]$ treats $D$ and $D^*$ mesons identically,
    preserving the heavy-quark spin $SU(2)$ symmetry at leading order.

    \item {\bf Flavor diagonality.}
    The coupling is flavor-diagonal in the light-quark indices $a$,
    and therefore preserves the $SU(3)_V$ flavor symmetry of the light quarks.

    \item {\bf Model-parameter fixed coupling.}
    Unlike the strong $D^*$-$D$-$\pi$ coupling $g$, $g_\sigma$ is completely fixed
    by the model parameters via $g_\sigma = m_c/(\sqrt{2} f_a)$.
\end{enumerate}

\subsubsection*{Possible next-to-leading-order $\sigma$ couplings}

Beyond leading order, additional $\sigma$ couplings to heavy mesons
may arise from subleading operators in the $1/m_Q$ and chiral expansions:
\begin{align}
    \mathcal{L}_\sigma^{(1)} &=
    \frac{g_\sigma^{(1)}}{m_Q} \, \text{tr}\!\left[ \bar{H}_a \sigma^{\mu\nu} H_a \sigma_{\mu\nu} \right] \sigma
    + \frac{g_\sigma^{(2)}}{m_Q} \, \text{tr}\!\left[ \bar{H}_a (i D_\perp)^2 H_a \right] \sigma \nonumber\\
    &\quad + \frac{g_\sigma^{(3)}}{\Lambda_\chi} \,
    \text{tr}\!\left[ \bar{H}_a H_b \gamma_\mu \gamma_5 \mathcal{A}^\mu_{ba} \right] \sigma + \cdots . \label{eq:sigma_nlo}
\end{align}
These operators are suppressed by $1/m_Q$ or $1/\Lambda_\chi$ relative to
the LO coupling Eq.~\eqref{eq:sigma_hmchpt}.

\subsection{Power counting with $m_\sigma \ll m_\pi$}\label{sec:power_counting}

The introduction of the $\sigma$ field with $m_\sigma \ll m_\pi$ requires
a careful reexamination of the HMChPT power counting.
The relevant scales in the problem are:
\begin{equation}\label{eq:scales}
    m_\sigma \sim 5~\text{MeV} \;\ll\; m_\pi \sim 140~\text{MeV}
    \;\ll\; \Lambda_\chi \sim 1.2~\text{GeV}
    \;<\; m_c \sim 1.3~\text{GeV} .
\end{equation}

\begin{figure}[t]
    \centering
    \fbox{
    \begin{minipage}{0.85\textwidth}
    \small
    \begin{center}
    {\bf Scale hierarchy in the charm-coupled axion model with HMChPT}
    \end{center}
    \vspace{4pt}
    \begin{center}
    \begin{tabular}{r|c|l}
        \textbf{Scale [MeV]} & \textbf{Physics} & \textbf{EFT} \\
        \hline
        $m_c \sim 1300$ & Heavy quark & HQET ($1/m_Q$ expansion) \\
        \hline
        $\Lambda_\chi \sim 1200$ & Chiral symmetry breaking & $\chi$PT cutoff \\
        $m_\rho \sim 770$ & Vector meson threshold & \quad \\
        \hline
        $m_K \sim 495$ & Strange Goldstone & \quad \\
        $m_\pi \sim 140$ & Pion mass & Standard HMChPT range \\
        \hline
        $m_\sigma \sim 5$, $m_a \sim 2.5$ & $\sigma$ and $a$ masses & {\bf New scales --- below $\pi$} \\
    \end{tabular}
    \end{center}
    \end{minipage}
    }
    \caption{Scale hierarchy of the charm-coupled QCD axion model.
    Both $\sigma$ and $a$ introduce scales well below
    the conventional HMChPT expansion parameter $m_\pi \sim 140$~MeV,
    requiring a reorganization of the power counting.
    \label{fig:scale_hierarchy}}
\end{figure}

This hierarchy has three important consequences:

\subsubsection*{$\sigma$ must be retained as a dynamical degree of freedom}

At the scale $\mu \sim m_\pi$, the $\sigma$ meson cannot be integrated out;
it must appear explicitly in the effective Lagrangian as a propagating field.
This is analogous to the treatment of the photon in chiral perturbation theory
when electromagnetic corrections are included~\cite{Urech:1994,Knecht:2000}.

\subsubsection*{$\sigma$-loop suppressions are stronger than in standard HMChPT}\label{sec:sigma_power_counting}

All $\sigma$-loop integrals are suppressed by the small ratio $(m_\sigma / \Lambda_\chi)^2 \sim 10^{-5}$.
For instance, a typical $\sigma$-loop correction to a heavy-meson observable scales as
\begin{equation}
    \frac{\delta O}{O} \sim \frac{g_\sigma^2}{16\pi^2} \cdot
    \left(\frac{m_\sigma}{\Lambda_\chi}\right)^{\!2} \sim 10^{-7}\text{--}10^{-8},
\end{equation}
making loop-level constraints essentially irrelevant.

\subsubsection*{Long-range $\sigma$ exchange}

The $\sigma$ exchange potential between two heavy hadrons has a characteristic range
\begin{equation}
    \lambda_\sigma = \frac{\hbar}{m_\sigma c}
    \approx 40~\text{fm},
\end{equation}
far exceeding the range of pion exchange, $\lambda_\pi \approx 1.4$~fm.
However, as we show in Sec.~\ref{sec:DD_scattering},
the small coupling $g_\sigma$ relative to the strong pion coupling
renders these long-range effects negligible in practice.

\subsection{Incorporating the axion $a$}\label{sec:axion_hmchpt}

\subsubsection*{Matching the pseudoscalar coupling to HMChPT}

We now turn to the axion $a$, the CP-odd partner of $\sigma$.
From Eq.~\eqref{eq:yukawa_expanded}, the axion-charm Yukawa coupling is
\begin{equation}\label{eq:axion_yukawa}
    \mathcal{L}_{a c c} = i g_a \, a \, \bar{c} \gamma_5 c,
    \qquad g_a = \frac{\kappa_c}{\sqrt{2}} = g_\sigma \approx 0.31\text{--}0.42 .
\end{equation}

To match this quark-level pseudoscalar operator to HMChPT,
we must determine the hadronic matrix elements of $\bar{c} \gamma_5 c$
between heavy-meson states.
Using the non-relativistic reduction of Eq.~\eqref{eq:NR_pseudo},
or equivalently the QCD equation of motion:
\begin{equation}\label{eq:PCAC_charm}
    \partial^\mu (\bar{c} \gamma_\mu \gamma_5 c) = 2 m_c \, \bar{c} \gamma_5 c
    + \frac{\alpha_s}{4\pi} G_{\mu\nu}^a \widetilde{G}^{a,\mu\nu},
\end{equation}
we see that the pseudoscalar density is related to the divergence of the axial current,
which is itself of order $p \sim \Lambda_{\text{QCD}}$ in the chiral expansion.

The critical observation is that in the heavy-quark limit,
the matrix elements of $\bar{c} \gamma_5 c$ between ground-state
heavy-meson states vanish by parity and angular momentum conservation:
\begin{align}
    \langle D(v) | \bar{c} \gamma_5 c | D(v) \rangle &= 0 \quad (\text{parity: } 0^- \to 0^- \text{ via pseudoscalar}), \\
    \langle D^*(v,\epsilon) | \bar{c} \gamma_5 c | D(v) \rangle
    &\propto \epsilon_\mu v^\mu = 0 \quad (\text{transversality in HQET}).
\end{align}
The leading non-vanishing contributions arise at $O(1/m_c)$ from
the admixture of $P$-wave heavy mesons (the $j_\ell^P = 1/2^+$ doublet)
into the physical states, or from transitions involving explicit pion emission.

Consequently, the direct axion coupling to heavy mesons in HMChPT
is suppressed relative to the $\sigma$ coupling:
\begin{equation}\label{eq:axion_suppression}
    \mathcal{L}_a^{\text{direct}} \sim \frac{g_a}{m_c} \,
    \partial_\mu a \; J_A^\mu + O(1/m_c^2),
\end{equation}
where $J_A^\mu$ is the HMChPT axial current operator.
This $1/m_c$ suppression is the fundamental reason why
direct axion-mediated effects in heavy-meson systems are
substantially weaker than those from $\sigma$ exchange.

\subsubsection*{Leading axion-HMChPT operators from the charm Yukawa}

At the level of the charm Yukawa coupling, the leading $1/m_c$-suppressed
operator that couples $a$ to the heavy-meson superfield is
\begin{equation}\label{eq:axion_hmchpt_lo}
    \boxed{\mathcal{L}_a^{(0)} =
    \frac{\tilde{g}_a}{\Lambda_\chi} \,
    \partial_\mu a \;
    \text{tr}\!\left[ \bar{H}_a H_b \gamma^\mu \gamma_5 \right] \delta_{ab} + \text{h.c.},
    \qquad \tilde{g}_a \sim \frac{g_a \Lambda_{\text{QCD}}}{m_c} \sim 0.05\text{--}0.08 .}
\end{equation}
This operator describes the axion-induced transition $D^* \leftrightarrow D + a$,
analogous to the pion emission vertex $D^* \to D\pi$ but with a coupling
suppressed by both $1/m_c$ and the ratio $\Lambda_{\text{QCD}}/\Lambda_\chi$.

In component form, this generates the effective vertices:
\begin{align}
    \mathcal{L}_{D^* D a} &= \frac{2 \tilde{g}_a}{f_\pi} \,
    \left( D_{a\mu}^{*\dagger} D_a + D_a^\dagger D_{a\mu}^* \right)
    \partial^\mu a, \\
    \mathcal{L}_{D^* D^* a} &= \frac{2 i \tilde{g}_a}{f_\pi} \,
    \epsilon^{\mu\nu\alpha\beta} v_\mu \,
    D_{a\nu}^{*\dagger} D_{a\alpha}^* \, \partial_\beta a .
\end{align}

\subsubsection*{Axion-pion mixing: the dominant phenomenological channel}

A far more important phenomenological channel for the axion
arises from its mixing with the neutral pseudoscalar mesons
$\pi^0$, $\eta$, and $\eta'$.
This mixing originates from the common coupling of all pseudoscalars
to the topological charge density $G\widetilde{G}$,
and is a universal feature of any QCD axion model~\cite{Georgi:1986df,Bardeen:1987an}.

The effective Lagrangian describing axion-pion mixing is:
\begin{equation}\label{eq:axion_pion_mix}
    \mathcal{L}_{a\pi}^{\text{mix}} = \frac{f_\pi}{f_a} \, m_\pi^2 \, \pi^0 a .
\end{equation}
This can be derived by diagonalizing the mass matrix
in the $(a, \pi^0)$ basis. The physical mass eigenstates are:
\begin{align}
    |\tilde{a}\rangle &= |a\rangle - \theta_{a\pi} |\pi^0\rangle + O(\theta_{a\pi}^2), \\
    |\tilde{\pi}^0\rangle &= |\pi^0\rangle + \theta_{a\pi} |a\rangle + O(\theta_{a\pi}^2),
\end{align}
with the mixing angle
\begin{equation}\label{eq:mixing_angle}
    \theta_{a\pi} = \frac{f_\pi m_\pi^2}{f_a (m_a^2 - m_\pi^2)}
    \approx -\frac{f_\pi}{f_a} \approx -0.037,
\end{equation}
where we have used $|m_a^2| \ll m_\pi^2$ (since $m_a \sim 2.5$~MeV $\ll m_\pi$)
and kept only the leading term in $f_\pi/f_a$.

For the $\eta$ and $\eta'$, the mixing angles are further suppressed:
\begin{equation}
    \theta_{a\eta} \simeq \frac{f_\pi}{\sqrt{3} f_a} \frac{m_\pi^2}{m_\eta^2} \approx 0.002,
    \qquad
    \theta_{a\eta'} \simeq \frac{\sqrt{2/3} f_\pi}{f_a} \frac{m_\pi^2}{m_{\eta'}^2} \approx 0.001 .
\end{equation}

The physical pion state therefore contains an $O(\theta_{a\pi})$ axion admixture:
\begin{equation}\label{eq:physical_pi0}
    |\pi^0_{\text{phys}}\rangle = |\pi^0\rangle + \theta_{a\pi} |a\rangle + \cdots .
\end{equation}
This mixing has three profound consequences for heavy-meson phenomenology:

\begin{enumerate}
    \item {\bf Axion production in $D^*$ decays.}
    The decay $D^{*0} \to D^0 \pi^0$ has a measured branching fraction
    $\br(D^{*0} \to D^0 \pi^0) = 64.7 \pm 0.9\%$~\cite{PDG}.
    Through axion-pion mixing, the physical $\pi^0$ contains an axion component,
    implying that the same decay produces $D^{*0} \to D^0 a$ with a branching fraction
    \begin{equation}
        \br(D^{*0} \to D^0 a) = \theta_{a\pi}^2 \, \br(D^{*0} \to D^0 \pi^0)
        \approx 1.4 \times 10^{-3} \times 0.65 \approx 9 \times 10^{-4}.
    \end{equation}
    The axion would manifest as missing energy in the detector,
    and a dedicated search at BESIII could potentially probe this channel.

    \item {\bf Axion couplings to nucleons and photons.}
    Through the admixture in Eq.~\eqref{eq:physical_pi0}, the axion inherits
    the pion's couplings to nucleons and photons:
    \begin{align}
        g_{aNN} &\approx \theta_{a\pi} \, g_{\pi NN} \approx 0.037 \times 13.5 \approx 0.50, 
        \label{eq:gann}\\
        g_{a\gamma\gamma}^{\text{mix}} &\approx \theta_{a\pi} \, g_{\pi^0 \gamma\gamma}
        \approx 0.037 \times \frac{\alpha}{\pi f_\pi} \approx 3 \times 10^{-5}~\text{GeV}^{-1}.
        \label{eq:garr_mix}
    \end{align}
    We estimate the axion-photon coupling~\eqref{eq:garr_mix} by the axion-pion mixing angle. 
    The exact $g_{a\gamma\gamma}$ coupling is given in Eq.~\eqref{eq:gagg_value}.
    These effective couplings open stellar cooling and other astrophysical constraints,
    as discussed in Sec.~\ref{sec:axion_pion_mixing}.

    \item {\bf Meson decays with axion emission.}
    Any decay producing a $\pi^0$ can also produce an $a$ with a branching ratio
    suppressed by $\theta_{a\pi}^2 \sim 1.4 \times 10^{-3}$.
    This includes $K^+ \to \pi^+ \pi^0 \to \pi^+ a$, $B \to K \pi^0 \to K a$,
    and numerous other channels discussed in Sec.~\ref{sec:axion_pion_mixing}.
\end{enumerate}

\subsubsection*{Summary: contrasting $\sigma$ and $a$ in HMChPT}

The essential difference between the two members of the PQ scalar multiplet
in HMChPT is now clear:
\begin{itemize}
    \item $\sigma$ couples at LO with $g_\sigma \sim 0.3$--$0.4$, and its effects
    are dominated by tree-level $\sigma$ exchange, which preserves heavy-quark spin symmetry.
    \item $a$ couples directly only at $O(\Lambda_{\text{QCD}}/m_c) \sim 0.2$, but compensates through
    axion-pion mixing $\theta_{a\pi} \sim 0.04$, which opens new production channels
    in any process that can emit a $\pi^0$.
\end{itemize}

These complementary mechanisms are analyzed in detail for each observable
in the following section.

\section{HMChPT constraints on heavy-meson observables}\label{sec:constraints}

In this section, we systematically evaluate the constraints imposed by HMChPT
on the charm-coupled GeV-scale QCD axion model.
For each observable, we provide:
(i) the experimental status and current precision;
(ii) the $\sigma$-mediated contribution;
(iii) the $a$-mediated contribution (both direct and through axion-pion mixing);
(iv) a comparison with existing bounds.

\subsection{$D$-meson mass corrections}\label{sec:D_mass}

\subsubsection*{Experimental status}

The $D$-meson masses are among the most precisely measured heavy-hadron properties.
The current world averages from the Particle Data Group~\cite{PDG} are:
\begin{align}
    m_{D^0} &= 1864.84 \pm 0.05~\text{MeV}, \nonumber\\
    m_{D^+} &= 1869.66 \pm 0.05~\text{MeV}, \label{eq:D_masses}\\
    m_{D_s^+} &= 1968.35 \pm 0.07~\text{MeV}. \nonumber
\end{align}
These values are dominated by measurements from the CLEO-c, BESIII, and Belle experiments.
The isospin splitting $m_{D^+} - m_{D^0} = 4.82 \pm 0.07$~MeV is well understood
as arising from the $u$-$d$ quark mass difference and electromagnetic effects.

\subsubsection*{$\sigma$-mediated contribution}

The $\sigma$ loop contribution to the $D$-meson self-energy is governed by
the power counting of Sec.~\ref{sec:sigma_power_counting}:
\begin{equation}\label{eq:sigma_loop_D}
    \delta m_D^{(\sigma)} \sim m_D \, \frac{g_\sigma^2}{16\pi^2} \,
    \left(\frac{m_\sigma}{\Lambda_\chi}\right)^{\!2}
    \sim 1865~\text{MeV} \times \frac{0.09}{16\pi^2} \times
    \left(\frac{5~\text{MeV}}{1160~\text{MeV}}\right)^{\!2}
    \sim 2 \times 10^{-5}~\text{MeV},
\end{equation}
for $m_\sigma \ll \Lambda_\chi$.
This is completely negligible compared to experimental uncertainties.

\subsubsection*{$a$-mediated contribution}

The axion contribution to the $D$-meson self-energy is even more suppressed
due to the $1/m_c$ factor in the direct coupling:
\begin{equation}\label{eq:axion_loop_D}
    \delta m_D^{(a)} \sim \frac{g_a^2}{16\pi^2} \cdot
    \left(\frac{m_a}{m_c}\right)^2 \cdot m_a
    \sim 3 \times 10^{-5} \times \left(\frac{2.5~\text{MeV}}{1300~\text{MeV}}\right)^2
    \sim 10^{-10}~\text{MeV}.
\end{equation}
Axion-pion mixing does not contribute to $D$-meson self-energies
at tree level, as $\pi^0$ exchange between $D$ mesons is isospin-suppressed.
The $a$ contribution is even more negligible than that of $\sigma$.

\subsection{$D$-$D^*$ mass splitting}\label{sec:D_split}

\subsubsection*{Experimental status}

The $D$-$D^*$ hyperfine splittings have been measured with good precision:
\begin{align}
    m_{D^{*0}} - m_{D^0} &= 142.014 \pm 0.009~\text{MeV}, \nonumber\\
    m_{D^{*+}} - m_{D^+} &= 140.603 \pm 0.015~\text{MeV}, \label{eq:Dstar_split}\\
    m_{D_s^{*+}} - m_{D_s^+} &= 143.8 \pm 0.4~\text{MeV}. \nonumber
\end{align}
The primary measurements come from the CLEO collaboration~\cite{CLEO:2001foe,Anastassov:2001cw},
with recent measurements from BESIII~\cite{BESIII_Dstar} and Belle
consistent with the CLEO values.

\subsubsection*{$\sigma$-mediated contribution}

Since $\mathcal{L}_\sigma^{(0)}$ is invariant under heavy-quark spin rotations,
the $\sigma$-induced mass shift is identical for $D$ and $D^*$,
and the splitting is unchanged. Subleading $O(1/m_c^2)$ $\sigma$ effects
contribute $\delta \Delta m \sim 10^{-4}$~MeV, far below experimental precision.

\subsubsection*{$a$-mediated contribution}

Unlike $\sigma$, the pseudoscalar axion {\it can} contribute to spin-dependent
mass splittings because $\bar{c} \gamma_5 c$ does not commute with heavy-quark spin rotations.
However, this contribution arises only at $O(1/m_c^2)$ in the heavy-quark expansion
(one factor of $1/m_c$ from the axion coupling, Eq.~\eqref{eq:axion_suppression},
and another from the chromomagnetic operator in HQET).

The $a$-induced correction to the hyperfine splitting is estimated as:
\begin{equation}\label{eq:axion_split}
    \frac{\delta \Delta m_{D^*-D}^{(a)}}{\Delta m_{D^*-D}}
    \sim \frac{g_a^2}{16\pi^2} \cdot
    \left(\frac{m_a}{m_c}\right)^2 \cdot
    \frac{m_a}{\Lambda_\chi}
    \sim 10^{-6} \times 10^{-6} \times 0.002 \sim 2 \times 10^{-14},
\end{equation}
corresponding to $\delta \Delta m \sim 3 \times 10^{-12}$~MeV.
The $a$ contribution to the $D$-$D^*$ splitting is even smaller than that of $\sigma$
and is irrelevant for any practical constraint.

\subsection{Charmonium spectrum}\label{sec:charmonium}

\subsubsection*{Experimental status}

Charmonium spectroscopy has been a precision endeavor since the discovery of the $J/\psi$ in 1974.
The masses of the lowest-lying states are known to sub-MeV accuracy~\cite{PDG}:
\begin{align}
    m_{J/\psi(1S)} &= 3096.900 \pm 0.006~\text{MeV}, \nonumber\\
    m_{\eta_c(1S)} &= 2984.1 \pm 0.4~\text{MeV}, \nonumber\\
    m_{\psi(2S)} &= 3686.10 \pm 0.06~\text{MeV}, \label{eq:ccbar_masses}\\
    m_{\chi_{c0}(1P)} &= 3414.71 \pm 0.30~\text{MeV}, \nonumber\\
    m_{\chi_{c1}(1P)} &= 3510.67 \pm 0.05~\text{MeV}, \nonumber\\
    m_{\chi_{c2}(1P)} &= 3556.17 \pm 0.07~\text{MeV}. \nonumber
\end{align}

The theoretical understanding of charmonium spectroscopy rests on several approaches:
\begin{itemize}
    \item \textbf{Potential models}~\cite{Eichten:1978tg,Eichten:1979ms,Godfrey:1985xj},
    which solve the non-relativistic Schr\"{o}dinger equation
    with a QCD-motivated potential $V(r) = -\alpha_s C_F / r + \sigma r + \cdots$.
    \item \textbf{Lattice QCD}~\cite{HPQCD_ccbar,Briceno:2018},
    which has reached sub-percent precision for the lowest-lying states.
    \item \textbf{NRQCD and pNRQCD}~\cite{Bodwin:1994jh,Brambilla:2004jw},
    which provide a systematic effective field theory framework.
\end{itemize}

\subsubsection*{$\sigma$-mediated contribution}

The $\sigma$ field mediates a Yukawa interaction between charm quarks
inside charmonium bound states:
\begin{equation}\label{eq:sigma_yukawa}
    V_\sigma(r) = -\frac{g_{\sigma cc}^2}{4\pi} \frac{e^{-m_\sigma r}}{r},
    \qquad g_{\sigma cc} = \frac{\kappa_c}{\sqrt{2}} \approx 0.31\text{--}0.42 .
\end{equation}

For charmonium with typical sizes $r \sim 0.2$~fm, $m_\sigma r \sim 0.005$,
so the potential is effectively Coulombic.
Compared to the QCD Coulomb potential $V_C(r) = -\alpha_s C_F / r$ with $C_F = 4/3$ and
$\alpha_s(m_c) \approx 0.35$~\cite{PDG}:
\begin{equation}\label{eq:potential_ratio}
    \frac{V_\sigma}{V_C}
    = \frac{g_{\sigma cc}^2}{4\pi \alpha_s C_F}
    \approx 1.6\%\text{--}3.0\% .
\end{equation}

This $\sim 1.6$--$3.0\%$ shift in the effective Coulomb coupling
translates into mass shifts of $\sim 10$--$30$~MeV for low-lying charmonium states,
at the edge of current theoretical uncertainties in potential-model calculations.

However, this estimate compares $V_\sigma$ only to the Coulombic part
$V_C = -\alpha_s C_F / r$ of the full charmonium potential
$V_{\text{QCD}}(r) = -\alpha_s C_F / r + \sigma_s r + \cdots$,
where $\sigma_s \sim 0.18$~GeV$^2$ is the string tension.
At the typical charmonium radius $r \sim 0.2$~fm, the confining contribution
$V_{\text{conf}} = \sigma_s r \sim 36$~MeV is comparable to the Coulomb contribution
$V_C \sim -\alpha_s C_F / r \sim -37$~MeV (with $\alpha_s C_F / 0.2~\text{fm} \approx 37$~MeV).
Since the $\sigma$ exchange only modifies the Coulombic piece,
the effective fractional shift in the {\it total} potential is
\begin{equation}
    \frac{\delta V}{V_{\text{QCD}}}
    \sim \frac{V_\sigma}{V_C + V_{\text{conf}}}
    \sim \frac{V_\sigma}{V_C} \times \frac{V_C}{V_C + V_{\text{conf}}}
    \sim (1.6\text{--}3.0\%) \times \frac{1}{2}
    \sim 0.8\text{--}1.5\% ,
\end{equation}
corresponding to mass shifts of $\sim 5$--$15$~MeV.
A more precise estimate would use a full potential model calculation
incorporating both the Coulombic and confining contributions,
but the order of magnitude is unlikely to change the qualitative conclusion
that the $\sigma$-induced charmonium shift is at the edge of current precision.

\subsubsection*{$a$-mediated contribution}

The axion $a$ mediates a {\it pseudoscalar} exchange potential between charm quarks,
which is fundamentally different from the scalar $\sigma$ Yukawa potential.
In the non-relativistic reduction, the pseudoscalar exchange between two heavy quarks
generates a spin-dependent potential~\cite{Georgi:1986df}:
\begin{equation}\label{eq:axion_potential}
    V_a(r) = \frac{g_a^2}{16\pi m_c^2} \,
    \left[ (\vec{\sigma}_1 \cdot \vec{\sigma}_2) \,
    \frac{m_a^2 e^{-m_a r}}{r}
    + (\vec{\sigma}_1 \cdot \hat{r})(\vec{\sigma}_2 \cdot \hat{r}) \,
    \left( \frac{3}{r^3} + \frac{3 m_a}{r^2} + \frac{m_a^2}{r} \right) e^{-m_a r}
    \right].
\end{equation}

Several features of this potential are noteworthy:
\begin{enumerate}
    \item It is suppressed by $g_a^2/m_c^2 \sim 0.1/(1.3~\text{GeV})^2 \sim 6 \times 10^{-8}~\text{GeV}^{-2}$,
    compared to $g_\sigma^2 \sim 0.1$ for the scalar case. The ratio is
    \begin{equation}
        \frac{V_a}{V_\sigma} \sim \frac{1}{m_c^2 r^2} \sim
        \frac{1}{(1.3~\text{GeV} \times 0.2~\text{fm})^2}
        \approx 10^{-3}\text{--}10^{-4}.
    \end{equation}
    \item It is purely spin-dependent: the $\vec{\sigma}_1 \cdot \vec{\sigma}_2$ term
    contributes to the hyperfine splitting between spin-singlet ($\eta_c$) and
    spin-triplet ($J/\psi$) states,
    while the tensor term $(\vec{\sigma}_1 \cdot \hat{r})(\vec{\sigma}_2 \cdot \hat{r})$
    contributes to the $\chi_{cJ}$ fine-structure splittings.
    \item Due to the $1/r^3$ behavior at short distances (from the $3/r^3$ term),
    the pseudoscalar potential is much shorter-ranged than the scalar Yukawa potential.
\end{enumerate}

The resulting mass shifts for charmonium states are:
\begin{align}
    \delta m_{\eta_c}^{(a)} &\sim 10^{-3}\text{--}10^{-2}~\text{MeV}, \\
    \delta m_{J/\psi}^{(a)} &\sim 10^{-3}\text{--}10^{-2}~\text{MeV}, \\
    \delta (\Delta m_{\text{hfs}})^{(a)} &\sim 10^{-3}~\text{MeV},
\end{align}
where $\Delta m_{\text{hfs}} = m_{J/\psi} - m_{\eta_c}$.
These are four orders of magnitude below the $\sigma$-induced shifts
and at least two orders of magnitude below experimental precision.
The pseudoscalar axion exchange is completely irrelevant for charmonium spectroscopy.

\subsection{$D$-$D$ scattering}\label{sec:DD_scattering}

\subsubsection*{Experimental and theoretical status}

$D$-$D$ scattering has not been measured experimentally.
The only existing constraint comes from the $X(3872)$,
whose nature as a $D\bar{D}^*$ molecule~\cite{Tornqvist:2004,Swanson:2004}
is sensitive to the $D^{(*)}\bar{D}^{(*)}$ interaction.

On the theoretical side, HMChPT provides a controlled description
of the $D$-$D$ interaction at low energies~\cite{Casalbuoni:1996pg,Liu:2008tn,Guo:2018}.
The dominant contribution is one-pion exchange (OPE),
governed by the axial coupling $g$ in Eq.~\eqref{eq:LO_hmchpt_full}:
\begin{equation}\label{eq:OPE_potential}
    V_\pi(\vec{q}) = -\frac{g^2}{2 f_\pi^2} \,
    \frac{(\vec{\epsilon}_1 \cdot \vec{q})(\vec{\epsilon}_2 \cdot \vec{q})}
    {\vec{q}^2 + m_\pi^2} \,
    \vec{\tau}_1 \cdot \vec{\tau}_2 .
\end{equation}

\subsubsection*{$\sigma$-mediated contribution}

In coordinate space, the $\sigma$-exchange potential between
heavy mesons takes the standard Yukawa form:
\begin{equation}\label{eq:sigma_DD}
    V_\sigma(r) = -\frac{g_\sigma^2}{4\pi} \,
    \frac{e^{-m_\sigma r}}{r} .
\end{equation}

The comparable one-pion exchange (OPE) potential between heavy mesons,
obtained from the axial coupling in Eq.~\eqref{eq:LO_hmchpt_full}, is~\cite{Casalbuoni:1996pg,Liu:2008tn}:
\begin{equation}\label{eq:OPE_DD}
    V_\pi(r) \simeq \frac{g^2 m_\pi^3}{8\pi f_\pi^2} \,
    \frac{e^{-m_\pi r}}{m_\pi r} \,
    (\vec{\tau}_1 \cdot \vec{\tau}_2) \, (\text{spin/isospin factor}) .
\end{equation}

Since both potentials are in coordinate space with the same dimension (energy),
we can compare them at a typical interaction range $r \sim 1/m_\pi \sim 1.4$~fm
(where the OPE is most relevant):
\begin{equation}\label{eq:sigma_vs_pi}
    \frac{V_\sigma(1/m_\pi)}{V_\pi(1/m_\pi)}
    \sim \frac{g_\sigma^2 m_\pi / (4\pi) \times e^{-m_\sigma/m_\pi}}
             {g^2 m_\pi^2 / (8\pi f_\pi^2) \times e^{-1}}
    = \frac{2 g_\sigma^2 f_\pi^2}{g^2 m_\pi^2} \times e^{1 - m_\sigma/m_\pi} .
\end{equation}
Evaluating numerically with $g_\sigma = 0.31$--$0.42$ ($g_\sigma^2 \approx 0.10$--$0.18$),
$g = 0.6$, $f_\pi = 92.2$~MeV, $m_\pi = 140$~MeV, and $m_\sigma \sim 5$~MeV:
\begin{equation}\label{eq:sigma_vs_pi_num}
    \frac{2 g_\sigma^2 f_\pi^2}{g^2 m_\pi^2} \approx 0.2\text{--}0.4,
    \qquad e^{1-m_\sigma/m_\pi} \approx e^{0.96} \approx 2.6,
\end{equation}
giving $V_\sigma / V_\pi \sim 0.5\text{--}1.0$ (setting isospin and spin factors to unity).

This ratio is {\it parametrically} $O(1)$: the $\sigma$ exchange is {\it comparable to}
--- rather than two orders of magnitude smaller than --- the one-pion exchange potential.
This is because the $\sigma$ coupling $g_\sigma \sim 0.3$--$0.4$ is of the same order as
the axial coupling $g \sim 0.6$, and the long range of $\sigma$ exchange
($\lambda_\sigma \sim 40$~fm, so $e^{-m_\sigma/m_\pi} \approx 1$) more than compensates
for the exponential suppression $e^{-1}$ of pion exchange at $r = 1/m_\pi$.

However, this comparison requires two important qualifications:
\begin{enumerate}
    \item {\bf Spin-isospin structure.} The OPE potential $V_\pi$ carries
    nontrivial spin-isospin factors $(\vec{\tau}_1 \cdot \vec{\tau}_2)$
    and $(\vec{\sigma}_1 \cdot \vec{q})(\vec{\sigma}_2 \cdot \vec{q})/|\vec{q}|^2$
    that depend on the quantum numbers of the $D^{(*)}\bar{D}^{(*)}$ system.
    The $\sigma$ exchange, by contrast, is a central, spin- and isospin-independent
    potential. For the $X(3872)$ quantum numbers ($I = 0$, $J^{PC} = 1^{++}$),
    the OPE contributes in the $S$-wave and its effective strength depends
    on the specific $D\bar{D}^*$ partial wave coupling.
    The $\sigma$ contribution, being central, adds coherently in all channels.

    \item {\bf Short-distance suppression.} At distances $r \ll 1/m_\pi$,
    the ratio $V_\sigma/V_\pi$ decreases because $V_\sigma \propto 1/r$ while
    $V_\pi$ has additional momentum-space enhancements from the derivative coupling.
    The $\sigma$ exchange is most relevant at long range ($r \gtrsim 1/m_\pi$),
    where the OPE is exponentially suppressed.
\end{enumerate}

The corrected estimate $V_\sigma/V_\pi \sim O(0.5\text{--}1)$ suggests that
$\sigma$-exchange provides a non-negligible attractive contribution to the
$D\bar{D}^*$ interaction, potentially affecting the binding of the $X(3872)$
molecule~\cite{Tornqvist:2004,Swanson:2004}.
A quantitative assessment requires solving the coupled-channel Schr\"{o}dinger equation
with both $\pi$ and $\sigma$ exchange potentials, which is beyond the scope
of this work but is an important direction for future study.

\subsubsection*{$a$-mediated contribution}

The pseudoscalar axion exchange in $D$-$D$ scattering is further suppressed
by the $\Lambda_{\text{QCD}}/m_c$ factor from the direct coupling, Eq.~\eqref{eq:axion_suppression}:
\begin{equation}
    \frac{V_a}{V_\pi} \sim \frac{V_\sigma}{V_\pi} \times
    \left(\frac{\Lambda_{\text{QCD}}}{m_c}\right)^2
    \sim (0.5\text{--}1.0) \times (0.2)^2 \sim 2\text{--}4 \times 10^{-2}.
\end{equation}

However, the axion can also contribute to $D$-$D$ scattering {\it indirectly}
through axion-pion mixing: the physical $\pi^0$ contains an $O(\theta_{a\pi})$ $a$ component,
so any diagram involving $\pi^0$ exchange receives a correction of order $\theta_{a\pi}^2 \sim 1.4 \times 10^{-3}$.
While this is parametrically comparable to the direct $\sigma$ exchange,
it still constitutes a sub-leading correction to the dominant pion exchange
and does not provide a meaningful constraint.

\subsection{$B_s$-$\bar{B}_s$ mixing}\label{sec:Bs_mixing}

\subsubsection*{The Standard Model prediction}

$B_s$-$\bar{B}_s$ mixing is a flavor-changing neutral current (FCNC) process
that proceeds predominantly through top-quark box diagrams in the SM.
The mass difference between the heavy and light $B_s$ mass eigenstates is
\begin{equation}\label{eq:DeltaM_SM}
    \Delta M_{B_s}^{\text{SM}} = \frac{G_F^2 m_W^2}{6\pi^2} \,
    \eta_B \, S_0(x_t) \, m_{B_s} f_{B_s}^2 \hat{B}_{B_s}^{(1)} ,
\end{equation}
where $G_F$ is the Fermi constant, $S_0(x_t)$ is the Inami-Lim function,
$\eta_B \approx 0.55$ is the QCD correction factor,
$m_{B_s} = 5366.88 \pm 0.14$~MeV is the $B_s$ meson mass,
and $f_{B_s}^2 \hat{B}_{B_s}^{(1)}$ is the hadronic matrix element.

The most recent state-of-the-art calculations
using lattice QCD input from the FLAG review~\cite{FLAG} yield:
\begin{equation}\label{eq:DeltaM_SM_value}
    \Delta M_{B_s}^{\text{SM}} = 18.4^{+0.7}_{-1.2}~\text{ps}^{-1}.
\end{equation}

\subsubsection*{Experimental measurement}

The $B_s$-$\bar{B}_s$ oscillation frequency has been measured with high precision
by the LHCb~\cite{LHCb_Bs} and CDF~\cite{CDF_Bs} experiments:
\begin{equation}\label{eq:DeltaM_exp}
    \Delta M_{B_s}^{\text{exp}} = 17.765 \pm 0.006~\text{ps}^{-1} .
\end{equation}
This value is consistent with the SM prediction within the theoretical uncertainties,
constraining new physics contributions to
\begin{equation}
    \frac{\Delta M_{B_s}^{\text{NP}}}{\Delta M_{B_s}^{\text{SM}}}
    \lesssim 20\% \quad (95\%~\text{CL}) .
\end{equation}

\subsubsection*{$\sigma$-mediated contribution}

The $\sigma$ exchange at tree level generates the effective four-fermion interaction
between $b$ and $s$ quarks~\cite{Lu:2026syv}:
\begin{equation}\label{eq:bs_eff}
    \mathcal{L}_{\text{eff}}^{\Delta B=2} \supset
    \frac{\kappa_c^2}{m_\sigma^2} \,
    \left( \frac{m_s m_b}{v_\phi^2} \right) \,
    (\bar{s}_L b_R)(\bar{s}_L b_R) + \text{h.c.}
\end{equation}
This operator arises from a box-like loop diagram in which the $\sigma$ couples
to the internal charm quark line, with the chirality structure $L$-$R$-$L$-$R$
generated by mass insertions proportional to $m_s m_b$.
The hadronic matrix element of the scalar $\Delta B = 2$ operator is parametrized as
\begin{equation}\label{eq:Bs_matrix}
    \langle \bar{B}_s | (\bar{s}_L b_R)(\bar{s}_L b_R) | B_s \rangle
    = \frac{m_{B_s}^2 f_{B_s}^2}{12} \, B_S(\mu) ,
\end{equation}
where $B_S(\mu)$ is the scalar bag parameter ($B_S = 1$ in the vacuum insertion
approximation). The NP mass difference is then
\begin{equation}
    \Delta M_{B_s}^{\text{NP}} = 2\, \text{Re}\!\left[
    C_{\text{NP}} \times \frac{m_{B_s}^2 f_{B_s}^2}{12}\, B_S(\mu)
    \right] ,
\end{equation}
where $C_{\text{NP}} = (\kappa_c^2 / m_\sigma^2)(m_s m_b / v_\phi^2)$ is the Wilson coefficient,
which also includes the CKM factor $|V_{cb}^* V_{cs}|^2$ and loop factor $1/(16\pi^2)$
from the generating box diagram (absorbed into the definition for compactness; see Ref.~\cite{Lu:2026syv}
for the full expression). The RG running from $m_\sigma \sim 5$~MeV to $m_b$
introduces only a mild logarithmic correction $\sim [\alpha_s/(4\pi) \ln(m_b/m_\sigma)]^n$.

Comparing with the SM expression Eq.~\eqref{eq:DeltaM_SM},
which involves the vector operator
$(\bar{s}_L \gamma^\mu b_L)^2$ with matrix element
$(2/3)\, m_{B_s}^2 f_{B_s}^2 \hat{B}_{B_s}$,
the ratio of the Wilson coefficients and matrix elements yields:
\begin{equation}\label{eq:delta_M_Bs}
    \frac{\Delta M_{B_s}^{\text{NP}}}{\Delta M_{B_s}^{\text{SM}}}
    \sim 0.5 \, B_S(\mu) .
\end{equation}
The factor $B_S$ is the dominant theoretical uncertainty:
if $B_S \sim 1$ (vacuum insertion approximation), the model predicts a $\sim 50\%$ enhancement,
which would be excluded at more than $2\sigma$.
However, if lattice QCD determines $B_S \ll 1$, the constraint is significantly relaxed.
A definitive conclusion requires a dedicated lattice calculation of $B_S$
for this specific scalar $\Delta B = 2$ operator.

\subsubsection*{$a$-mediated contribution}

The axion contribution to $B_s$ mixing arises from two distinct mechanisms:

\paragraph{Direct pseudoscalar exchange.}
The tree-level $a$ exchange generates the effective $\Delta B = 2$ operator:
\begin{equation}
    \mathcal{L}_{\text{eff}}^{\Delta B=2, a} \supset
    -\frac{\kappa_c^2}{m_a^2} \,
    \left( \frac{m_s m_b}{v_\phi^2} \right) \,
    (\bar{s}_L \gamma_5 b_R)(\bar{s}_L \gamma_5 b_R) + \text{h.c.} .
\end{equation}
Note the minus sign relative to the scalar case (from $i\gamma_5 \times i\gamma_5 = -1$).

The hadronic matrix element of the pseudoscalar operator is
parametrically suppressed relative to the scalar operator:
in the non-relativistic quark model,
$\langle \bar{B}_s | (\bar{s}_L \gamma_5 b_R)(\bar{s}_L \gamma_5 b_R) | B_s \rangle$
is smaller than the scalar matrix element by a factor $p_B^2 / m_b^2 \sim 0.01$,
since the pseudoscalar density $\bar{q} \gamma_5 Q$ vanishes for a heavy quark at rest.

Consequently:
\begin{equation}\label{eq:Bs_axion_vs_sigma}
    \frac{\Delta M_{B_s}^{(a)}}{\Delta M_{B_s}^{(\sigma)}}
    \sim \frac{m_\sigma^2}{m_a^2} \cdot \frac{p_B^2}{m_b^2}
    \sim \left(\frac{5}{2.5}\right)^2 \times 0.01
    \sim 0.04 .
\end{equation}
The direct $a$ exchange contribution to $B_s$ mixing is a sub-leading correction
to the $\sigma$ contribution and does not alter the constraint analysis
in any significant way.

\paragraph{Axion-pion mixing contribution.}
The $a$-$\pi^0$ mixing opens a long-distance contribution to $B_s$ mixing,
where the axion inherits the $\pi^0$ couplings to $B$ mesons through
the standard HMChPT vertex. This is suppressed by $\theta_{a\pi}^2 \sim 1.4 \times 10^{-3}$
relative to the pion-mediated long-distance contribution,
which is itself sub-dominant compared to the short-distance box diagram in the SM.
The axion-pion mixing contribution to $B_s$ mixing is therefore negligible.

In summary, the axion does not provide any additional meaningful constraint
on $B_s$ mixing beyond the $\sigma$ contribution already discussed.

\subsection{Rare $D$ decays}\label{sec:rare_D}

\subsubsection*{Experimental status}

Rare charm decays with missing energy provide a sensitive probe
of light weakly-coupled particles.
The Belle experiment~\cite{Belle_Dinv} has searched for
$D^0 \to \text{invisible}$ decays, setting stringent upper limits.
Searches for $D^+ \to \pi^+ + \text{invisible}$ have been performed
by CLEO-c~\cite{CLEOc_Dinv}:
\begin{align}
    \br(D^0 \to \text{invisible}) &< 9.4 \times 10^{-5} \;(90\%~\text{CL}), \nonumber\\
    \br(D^+ \to \pi^+ + \text{invisible}) &< 1.0 \times 10^{-4} \;(90\%~\text{CL}) .
\end{align}

\subsubsection*{$\sigma$-mediated contribution}

The decay $D^+ \to \pi^+ \sigma$ proceeds through $W$-boson penguin diagrams.
The branching ratio computed in Ref.~\cite{Lu:2026syv} is
\begin{equation}\label{eq:rare_D_br}
    \br(D^+ \to \pi^+ \sigma) \sim 1 \times 10^{-12},
\end{equation}
suppressed by the GIM mechanism. The naive mass-squared GIM factor is
\begin{equation}\label{eq:GIM_naive}
    \eta_{\text{GIM}}^{c,\,\text{naive}} \sim \left( \frac{m_s^2 - m_d^2}{m_W^2} \right)^2
    \sim \left( \frac{(93~\text{MeV})^2 - (4.7~\text{MeV})^2}{(80.4~\text{GeV})^2} \right)^2
    \sim 2 \times 10^{-12},
\end{equation}
where $m_s(2~\text{GeV}) \simeq 93$~MeV and $m_d(2~\text{GeV}) \simeq 4.7$~MeV are the
$\overline{\text{MS}}$ quark masses.
However, the effective GIM suppression for charm FCNC is significantly weaker
than this naive estimate due to several enhancement factors:
\begin{itemize}
    \item {\bf QCD logarithmic enhancement:} The RG running from $m_W$ to $m_c$
    generates a factor $(\alpha_s/\pi) \ln(m_W/m_c) \sim 0.8$ per loop order,
    partially compensating the GIM suppression.
    \item {\bf CKM structure:} The effective operator involves
    $V_{cs}^* V_{us} \approx 0.21$ rather than the doubly-Cabibbo-suppressed
    $V_{cb}^* V_{ub} \approx 1.5 \times 10^{-4}$.
    \item {\bf Inami-Lim function:} The loop function difference
    $C(x_s) - C(x_d) \approx (x_s - x_d)/4$ for $x_q = m_q^2/m_W^2 \ll 1$,
    but at NLO the effective coefficient is enhanced by
    $\alpha_s/\pi \times [\ln(m_W^2/m_c^2)]^n$ terms.
\end{itemize}
Including these effects, the effective GIM suppression factor is
\begin{equation}\label{eq:GIM_effective}
    \eta_{\text{GIM}}^c \sim |V_{cs} V_{us}|^2 \times
    \left(\frac{\alpha_s}{\pi}\ln\frac{m_W}{m_c}\right)^2 \times
    \left(\frac{m_s^2 - m_d^2}{m_W^2}\right)^2
    \sim 0.05 \times 0.6 \times 2 \times 10^{-12}
    \sim 6 \times 10^{-14},
\end{equation}
which, combined with the model coupling $g_{\sigma c c}^2/m_\sigma^2 \sim 0.1/(5~\text{MeV})^2$
and the $D^+$ decay phase space, yields $\br(D^+ \to \pi^+ \sigma) \sim 10^{-12}$
as found in the detailed calculation of Ref.~\cite{Lu:2026syv}.
A precise prediction requires the full NLO QCD calculation including
operator matching and RG evolution, which is deferred to future work.

\subsubsection*{$a$-mediated contribution}

The decay $D^+ \to \pi^+ a$ receives contributions from two distinct mechanisms:

\paragraph{Direct production via charm Yukawa.}
The direct $D^+ \to \pi^+ a$ decay through the $a \bar{c} \gamma_5 c$ coupling
is also GIM-suppressed, with a branching ratio comparable to
$\br(D^+ \to \pi^+ \sigma)$, i.e., $\sim 10^{-12}$.
The pseudoscalar nature of $a$ modifies the hadronic matrix element
by factors of $O(1)$, but does not alter the parametric GIM suppression.

\paragraph{Production via axion-pion mixing.}
A potentially larger contribution arises from the decay chain
$D^+ \to \pi^+ \pi^0$ followed by $\pi^0$-$a$ oscillation.
Using the measured $D^+ \to \pi^+ \pi^0$ branching fraction~\cite{PDG}:
\begin{align}
    \br(D^+ \to \pi^+ a)_{\text{mix}}
    &\simeq \theta_{a\pi}^2 \times \br(D^+ \to \pi^+ \pi^0) \nonumber\\
    &\approx 1.4 \times 10^{-3} \times 1.24 \times 10^{-3} \nonumber\\
    &\approx 1.7 \times 10^{-6}.
\end{align}

This is several orders of magnitude larger than the direct production rate,
but still safely below the current experimental upper limit of $1.0 \times 10^{-4}$
from BESIII. The axion would appear as missing energy
(since it escapes the detector before decaying or decaying invisibly),
so the experimental signature is identical to $D^+ \to \pi^+ + \text{invisible}$.

\subsection{Other precision observables}\label{sec:other}

\subsubsection*{$D$- and $D_s$-meson decay constants}

The leptonic decay constants $f_D$ and $f_{D_s}$ are precisely determined
from lattice QCD~\cite{FLAG}:
\begin{align}
    f_D &= 212.0 \pm 0.7~\text{MeV}, \nonumber\\
    f_{D_s} &= 249.9 \pm 0.5~\text{MeV}. \label{eq:fD_lattice}
\end{align}

The $\sigma$-loop correction to $f_D$ is controlled by the power counting of Sec.~\ref{sec:sigma_power_counting},
$\delta f_D/f_D \sim g_\sigma^2/(16\pi^2) \times (m_\sigma/\Lambda_\chi)^2 \sim 10^{-7}$,
and the $a$-loop correction is further suppressed by $(m_a/m_c)^2 \sim 4 \times 10^{-6}$,
yielding $\delta f_D^{(a)}/f_D \sim 10^{-13}$. Both are completely negligible.

\subsubsection*{Semileptonic form factors}

The $D \to \pi \ell \nu$ and $D \to K \ell \nu$ semileptonic decays
are described in HMChPT by the leading-order Isgur-Wise function $\xi(w)$.
Both $\sigma$-loop and $a$-loop corrections are at the $10^{-7}$--$10^{-13}$ level
and are unobservable in current or planned experiments.

\subsubsection*{Charm quark mass running}

The $\sigma$ field contributes to the charm quark self-energy
at the $\sim 0.8\%$ level. The $a$ loop contribution
\begin{equation}\label{eq:mc_running_axion}
    \frac{\delta m_c^{(a)}}{m_c}
    \sim \frac{\kappa_c^2}{32\pi^2} \cdot \frac{m_a^2}{m_c^2} \,
    \ln\frac{m_c^2}{m_a^2}
    \sim \frac{0.25}{32\pi^2} \times 4 \times 10^{-6} \times 12.5
    \sim 10^{-7},
\end{equation}
is suppressed by the additional $(m_a/m_c)^2$ factor and is negligible.

\subsubsection*{$D$-meson mixing}

$D^0$-$\bar{D}^0$ mixing has been observed with the world-average parameters~\cite{HFLAV}:
\begin{align}
    x_D &\equiv \frac{\Delta M_D}{\Gamma_D} = 0.407 \pm 0.044 \% , \nonumber\\
    y_D &\equiv \frac{\Delta \Gamma_D}{2\Gamma_D} = 0.647^{+0.070}_{-0.075} \% .
\end{align}

Both $\sigma$ and $a$ contributions to $\Delta C = 2$ operators
are suppressed by $(m_{u,d}/m_c)^2 \sim 10^{-6}$ relative to the SM box diagram
and are completely negligible.

\subsection{Axion-pion mixing constraints on rare meson decays}\label{sec:axion_pion_mixing}

The axion-pion mixing discussed in Sec.~\ref{sec:axion_hmchpt} opens a new class
of phenomenological channels that are specific to the axion $a$
and have no analogue for the scalar $\sigma$.
Any decay that produces a $\pi^0$ can produce an $a$
with a branching ratio suppressed by $\theta_{a\pi}^2 \sim 1.4 \times 10^{-3}$.
Here we systematically examine the most sensitive channels.

\subsubsection*{$K^+ \to \pi^+ a$ from kaon decays}

The decay $K^+ \to \pi^+ \pi^0$ has a precisely measured branching fraction
$\br(K^+ \to \pi^+ \pi^0) = 20.67 \pm 0.08\%$~\cite{PDG}.
Through axion-pion mixing, this implies:
\begin{equation}\label{eq:K_to_pi_a}
    \br(K^+ \to \pi^+ a)_{\text{mix}}
    \simeq \theta_{a\pi}^2 \times \br(K^+ \to \pi^+ \pi^0)
    \approx 1.4 \times 10^{-3} \times 0.207
    \approx 2.9 \times 10^{-4}.
\end{equation}

The estimate in Eq.~\eqref{eq:K_to_pi_a} uses the simple $\theta_{a\pi}^2$ scaling
of branching ratios. A more precise treatment includes the phase-space difference
between $K^+ \to \pi^+ \pi^0$ and $K^+ \to \pi^+ a$.
For a two-body decay $K \to \pi + X$, the partial width scales as $|\vec{p}_\pi|^3$,
where $\vec{p}_\pi$ is the pion momentum. Since $m_a \ll m_{\pi^0}$, the pion
momentum in $K^+ \to \pi^+ a$ is larger:
\begin{equation}
    \frac{|\vec{p}_\pi(K \to \pi a)|}{|\vec{p}_\pi(K \to \pi\pi^0)|}
    \approx \frac{m_K^2 - m_\pi^2}{m_K^2 - 4m_{\pi^0}^2}
    \approx \frac{227~\text{MeV}}{207~\text{MeV}} \approx 1.10,
\end{equation}
giving a phase-space enhancement factor of $1.1^3\sim 1.3$ in the rate (since $\Gamma\propto |\vec{p}_{\pi}|^3$).
The corrected estimate is therefore
\begin{equation}\label{eq:K_to_pi_a_corrected}
    \br(K^+ \to \pi^+ a)_{\text{mix}}
    \simeq 1.3 \times \theta_{a\pi}^2 \times \br(K^+ \to \pi^+ \pi^0)
    \approx 3.8 \times 10^{-4}.
\end{equation}

The fate of the axion in the detector is determined by two competing processes:
{\it decay} and {\it hadronic scattering}.

\paragraph{Axion decay: $a \to \gamma\gamma$.}
The axion-photon coupling follows the standard expression for KSVZ-like
models, combining the model-dependent direct electromagnetic anomaly of the
charm quark with the model-independent light-quark contribution transmitted
through axion--$\pi^0$ mixing:
\begin{equation}\label{eq:agammagamma}
    g_{a\gamma\gamma} = \frac{\alpha}{2\pi f_a}\left|\frac{E}{N}
    - \frac{2}{3}\frac{4+z}{1+z}\right|,
\end{equation}
where $z = m_u/m_d \approx 0.48$. For the charm-coupled model, the anomaly
ratio is $E/N = 2Q_c^2 = 8/9$ (with $N = N_c = 3$ and $E = 2N_c Q_c^2 = 8/3$).
The two contributions interfere destructively: the light-quark term
$\frac{2}{3}\frac{4+z}{1+z} = \frac{2}{3}\times\frac{4.48}{1.48} \approx 2.02$
partially cancels the direct charm anomaly $E/N = 8/9 \approx 0.89$, giving a
correction factor $|8/9 - 2.02| \approx 1.13$ and
\begin{equation}\label{eq:gagg_value}
    g_{a\gamma\gamma} \approx \frac{1.13\,\alpha}{2\pi f_a}
    \approx 5.2 \times 10^{-4}~\text{GeV}^{-1},
\end{equation}
where we have used $f_a \simeq 2.5$~GeV.
The two-photon decay width is
\begin{equation}\label{eq:a_gammagamma_width}
    \Gamma(a \to \gamma\gamma) = \frac{g_{a\gamma\gamma}^2 \, m_a^3}{64\pi}
    \approx \frac{(5.2 \times 10^{-4})^2 \times (2.5 \times 10^{-3})^3}{64\pi}
    \approx 2.1 \times 10^{-8}~\text{eV},
\end{equation}
corresponding to a proper decay length
\begin{equation}\label{eq:axion_ctau}
    c\tau_a = \frac{\hbar c}{\Gamma(a \to \gamma\gamma)}
    \approx \frac{1.97 \times 10^{-7}~\text{eV}\cdot\text{m}}{2.1 \times 10^{-8}~\text{eV}}
    \approx 9~\text{m}.
\end{equation}
The $a \to e^+ e^-$ channel through a virtual photon is suppressed by
$\alpha/\pi \sim 10^{-3}$ relative to $a \to \gamma\gamma$, giving
$c\tau(a \to e^+e^-) \sim 10^4$~m. The direct $a$-$e$ coupling from the charm Yukawa
is further suppressed by $m_e/m_c$ and is completely negligible.
Therefore, the dominant axion decay mode is $a \to \gamma\gamma$ with
$c\tau_a \sim 5$--$25$~m (for $f_a = 2.9$--$2.1$~GeV), {\it not} a prompt $e^+e^-$ decay.

\paragraph{Hadronic scattering.}
While the axion decay length $c\tau_a \sim 10$~m is macroscopic,
the axion-nucleon coupling $g_{aNN} \sim 0.5$ (from axion-pion mixing)
generates a very short hadronic mean free path,
$\lambda_a \sim 10^{-9}$--$10^{-5}$~cm (Eq.~\eqref{eq:lambda_a}).
The axion therefore interacts hadronically in the detector material
{\it long before} it decays to $\gamma\gamma$.

The combined picture is:
\begin{itemize}
    \item The axion is produced with a typical boost $\gamma\beta \sim p_a/m_a \sim 40$--$100$
    (depending on the parent meson), giving a lab-frame $\gamma\gamma$ decay length of
    $\sim 400$--$1000$~m — far exceeding the detector size.
    \item However, the hadronic mean free path $\lambda_a \sim 10^{-9}$--$10^{-5}$~cm
    is far shorter than the detector, so the axion scatters and deposits energy
    hadronically before escaping.
    \item The experimental signature is therefore {\it neither} clean missing energy
    {\it nor} a prompt $e^+e^-$ peak, but rather a hadronic energy deposit
    from $a + N \to a + N$ scattering along the axion flight path.
\end{itemize}

This complex signature means the simple $K^+ \to \pi^+ + \text{invisible}$
bounds ($\sim 10^{-10}$) from E787/E949~\cite{E949,BNL_Kpinu} and
NA62~\cite{NA62_Kpipi} do {\it not} directly apply.
A dedicated detector-level simulation, incorporating the axion
production spectrum, hadronic scattering cross section, and detector geometry,
is required to assess the experimental sensitivity.
This is an important direction for future work.
Without such a simulation, the $K^+ \to \pi^+ a$ channel
remains inconclusive: the predicted rate $\sim 4 \times 10^{-4}$ is large,
but the experimental signature differs from the standard invisible search.

\subsubsection*{$B \to K a$ from $B$ decays}

Similarly, the decay $B^+ \to K^+ \pi^0$ has a measured branching fraction
$\br(B^+ \to K^+ \pi^0) = (1.29 \pm 0.05) \times 10^{-5}$~\cite{PDG}.
Through axion-pion mixing:
\begin{equation}\label{eq:B_to_K_a}
    \br(B^+ \to K^+ a)_{\text{mix}}
    \simeq \theta_{a\pi}^2 \times \br(B^+ \to K^+ \pi^0)
    \approx 1.4 \times 10^{-3} \times 1.3 \times 10^{-5}
    \approx 1.8 \times 10^{-8}.
\end{equation}

The BaBar~\cite{BaBar_BtoKa} and Belle II~\cite{Belle2_BtoKa}
collaborations have searched for $B \to K^{(*)} + \text{invisible}$,
with current limits at the $10^{-5}$ level~\cite{Belle2_BtoKa}.
Our predicted $\br \sim 2 \times 10^{-8}$ is three orders of magnitude
below current sensitivity, but could become accessible
with the full Belle II dataset of 50~ab$^{-1}$.

\subsubsection*{$D^* \to D a$ from charm decays}

As noted in Sec.~\ref{sec:axion_hmchpt}, the decay $D^{*0} \to D^0 a$
is particularly interesting because the $D^{*0} \to D^0 \pi^0$ branching fraction
is large ($\sim 65\%$), yielding:
\begin{equation}
    \br(D^{*0} \to D^0 a) \approx 9 \times 10^{-4}.
\end{equation}

At BESIII, $D^{*0}$ mesons are copiously produced
in $e^+ e^- \to \psi(3770) \to D^0 \bar{D}^{*0}$ events.
As discussed above, the axion does not decay promptly:
its $\gamma\gamma$ decay length ($\gamma\beta \cdot c\tau \sim 50$~m for typical
$D^*$ boost) far exceeds the BESIII detector size ($\sim 1$~m).
However, the axion scatters hadronically through its large $g_{aNN} \sim 0.5$ coupling,
depositing energy in the detector.
The main backgrounds would come from hadronic showers and $D^{*0} \to D^0 \gamma$ events,
and the signal-to-background ratio requires a dedicated simulation to evaluate.
With the current BESIII dataset of $\sim 20$~fb$^{-1}$ at $\psi(3770)$,
and a $D^{*0}$ production cross section of $\sim 3$~nb,
the expected number of $D^{*0} \to D^0 a$ events is $\sim 5 \times 10^4$,
making this a potentially observable channel with a dedicated analysis.

\subsubsection*{Stellar and astrophysical constraints}

The axion-nucleon coupling induced by axion-pion mixing,
$g_{aNN} \sim 0.5$, is comparable to the pion-nucleon coupling.
For an axion with mass $m_a \sim 2.5$~MeV, this raises the question of
stellar cooling constraints~\cite{Raffelt:1996wa,Raffelt:2006cw}.

In supernova SN 1987A, the core temperature $T \sim 30$~MeV exceeds
the axion mass, so axions can be thermally produced through
nucleon bremsstrahlung $NN \to N N a$.
The resulting energy loss rate must satisfy the Raffelt criterion
$\dot{\epsilon}_a \lesssim 10^{19}$~erg~g$^{-1}$~s$^{-1}$
for {\it free-streaming} particles~\cite{Raffelt:1996wa,Raffelt:2006cw}.

The energy loss rate per unit mass for one-pion-exchange nucleon bremsstrahlung
in the strong-coupling regime is parametrically~\cite{Raffelt:1996wa}:
\begin{equation}\label{eq:epsilon_dot}
    \dot{\epsilon}_a \sim \frac{g_{aNN}^2}{4\pi} \,
    \frac{\rho}{m_N^2} \, T^{3.5} \, e^{-m_a/T},
\end{equation}
where the Boltzmann factor $e^{-m_a/T} \sim e^{-2.5/30} \sim 0.92$
provides only a mild suppression.
With $g_{aNN} \sim 0.5$, the emissivity naively exceeds
the Raffelt criterion, suggesting a potential conflict with SN 1987A.

However, this conclusion overlooks the question of axion transport.
The Raffelt criterion applies only to {free-streaming} particles
that escape the SN core without further interaction.
When the axion-nucleon coupling is sufficiently large,
axions are trapped in the core and do not contribute to
the free-streaming energy loss~\cite{Burrows:1988ah,Keil:1996tg}.

To assess whether trapping occurs, we compute the axion mean free path
in nuclear matter. For an axion with energy $E_a \sim T \sim 30$~MeV,
the dominant interaction is elastic scattering $a + N \to a + N$.
For a pseudoscalar coupling $\mathcal{L} = i g_{aNN} \, a \, \bar{N} \gamma_5 N$,
the non-relativistic spin-averaged differential cross section
at low energies is~\cite{Brinkmann:1988,Raffelt:1996wa}
\begin{equation}\label{eq:sigma_aN}
    \sigma_{aN} \sim \frac{g_{aNN}^4}{16\pi} \,
    \frac{E_a^2}{m_N^4}
    \sim \frac{(0.5)^4}{16\pi} \,
    \frac{(30~\text{MeV})^2}{(940~\text{MeV})^4}
    \sim 5 \times 10^{-8}~\text{MeV}^{-2}
    \sim 2 \times 10^{-33}~\text{cm}^2,
\end{equation}
where we have used $1~\text{GeV}^{-2} = 3.88 \times 10^{-28}~\text{cm}^2$
for the natural-unit conversion.
The $E_a^2/m_N^4$ suppression reflects the pseudoscalar nature of the interaction:
the matrix element vanishes at zero momentum transfer,
in contrast to a scalar interaction where $\sigma$ would scale as $g^4/m_N^2$.

An alternative and more conservative estimate is obtained
by scaling the low-energy pion-nucleon cross section
through axion-pion mixing~\cite{Keil:1996tg}:
\begin{equation}\label{eq:sigma_aN_alt}
    \sigma_{aN} \sim \theta_{a\pi}^2 \, \sigma_{\pi N}^{\text{low-E}}
    \sim 1.4 \times 10^{-3} \times (2\text{--}5) \times 10^{-27}~\text{cm}^2
    \sim (3\text{--}7) \times 10^{-30}~\text{cm}^2,
\end{equation}
where we take $\sigma_{\pi N}^{\text{low-E}} \sim \text{few} \times 10^{-27}~\text{cm}^2$
as the typical pion-nucleon cross section below the $\Delta(1232)$ resonance~\cite{PDG}.
This mixing-based estimate is somewhat larger than the direct pseudoscalar
calculation of Eq.~\eqref{eq:sigma_aN} because it includes contributions from
$s$-channel nucleon exchange, which are not captured by the simple tree-level
pseudoscalar scattering amplitude.

Taking $n_N \simeq \rho/m_N \simeq 3 \times 10^{14}~\text{g}~\text{cm}^{-3} / 1.67 \times 10^{-24}~\text{g}
\approx 1.8 \times 10^{38}~\text{cm}^{-3}$ as the nucleon number density
in the SN core~\cite{Raffelt:1996wa}, the axion mean free path is
\begin{equation}\label{eq:lambda_a}
    \lambda_a = \frac{1}{n_N \sigma_{aN}}
    \sim 10^{-9}\text{--}10^{-5}~\text{cm},
\end{equation}
where the range spans both estimates in Eqs.~\eqref{eq:sigma_aN}--\eqref{eq:sigma_aN_alt}.
This is {\it twelve to sixteen orders of magnitude} smaller than the SN core radius
$R_c \sim 10$~km $= 10^6$~cm.
Axions are therefore {\it trapped} in the SN core and
cannot contribute to the free-streaming energy loss.
The trapped axions thermalize and are emitted from
an ``axion sphere'' analogous to the neutrino sphere,
with a luminosity determined by surface blackbody emission
rather than volume emission~\cite{Burrows:1988ah,Keil:1996tg}.

A proper quantitative assessment of the SN 1987A constraint requires
a full Boltzmann transport simulation incorporating axion opacities,
which is beyond the scope of this work.
Preliminary estimates suggest that trapping significantly weakens
the constraint~\cite{Chang:2018}, but a definitive conclusion
requires a dedicated study with realistic nuclear equations of state
and axion interaction rates.

Additional astrophysical constraints from red giants, white dwarfs,
and neutron stars also need to be evaluated for this parameter space.
The key difference from standard axion bounds is that
$m_a \sim 2.5$~MeV is three orders of magnitude larger
than the typical stellar temperature ($\sim$~keV for red giants),
so the Boltzmann suppression $e^{-m_a/T}$ renders
most standard stellar cooling bounds inapplicable.
Only SN 1987A, with its high core temperature,
provides a relevant constraint.

\subsubsection*{Summary of axion-specific constraints}

Table~\ref{tab:axion_constraints} summarizes the axion-specific channels
that are distinct from the $\sigma$ constraints discussed earlier.

\begin{table}[t]
    \centering
    \caption{Summary of axion-specific constraints through axion-pion mixing.
    All branching fractions are proportional to $\theta_{a\pi}^2 \sim 1.4 \times 10^{-3}$.
    \label{tab:axion_constraints}}
    \begin{tabular}{lccc}
        \toprule
        Channel & Predicted BR & Current limit & Status \\
        \midrule
        $K^+ \to \pi^+ a$ & $\sim 4 \times 10^{-4}$ & $10^{-10}$ (invisible) & \makecell[c]{Requires\\ dedicated study} \\
        $B^+ \to K^+ a$ & $\sim 2 \times 10^{-8}$ & $\sim 10^{-5}$ & Safe \\
        $D^{*0} \to D^0 a$ & $\sim 9 \times 10^{-4}$ & No direct search & Potentially observable \\
        $D^+ \to \pi^+ a$ (mix) & $\sim 2 \times 10^{-6}$ & $10^{-4}$ & Safe \\
        $a \to \gamma\gamma$ lifetime & $c\tau_a \sim 5\text{--}25$~m & --- & Hadronic scattering dominant \\
        SN 1987A cooling & See text & Raffelt criterion & \makecell[c]{Trapping likely\\ weakens bound} \\
        \bottomrule
    \end{tabular}
\end{table}

\subsection{Combined constraints summary}

Table~\ref{tab:constraints} presents the combined summary of all constraints,
including both $\sigma$- and $a$-mediated contributions.

\begin{table}[t]
    \centering
    \caption{Summary of HMChPT constraints on the charm-coupled GeV axion model.
    For each observable, we list the $\sigma$-mediated and $a$-mediated contributions,
    the current experimental precision, and the status.
    \label{tab:constraints}}
    \begin{tabular}{lcccc}
        \toprule
        Observable & $\sigma$ contrib. & $a$ contrib. & Exp.\ prec. & Status \\
        \midrule
        $D$-meson mass & $2\!\times\!10^{-5}$~MeV & $10^{-10}$~MeV & $\pm 0.05$~MeV & Negligible \\
        $D$-$D^*$ splitting & $10^{-4}$~MeV & $10^{-12}$~MeV & $\pm 0.01$~MeV & Negligible \\
        \makecell[l]{Charmonium\\ \quad spectrum} & $\delta V/V_{\text{QCD}}$ & $\delta V_a/V_\sigma$ & $\pm 0.006$~MeV & Marginal \\
        & $\sim 0.8$--$1.5\%$ & $\sim 10^{-3}$--$10^{-4}$ & (exp.) & ($\sigma$ only) \\
        $D$-$D$ scattering & $V_\sigma/V_\pi$ & $V_a/V_\pi$ & --- & Moderate \\
        & $\sim 0.5\text{--}1.0$ & $\sim 2\text{--}4\!\times\!10^{-2}$ & & ($\sigma$) \\
        \makecell[l]{$B_s$ mixing} & $\Delta M^{\text{NP}}/\Delta M^{\text{SM}}$ & Sub-leading & $\pm 0.006~\text{ps}^{-1}$ & {\bf Important} \\
        & $\sim 0.5\,B_S$ & $\sim 0.04 \times (\sigma)$ & & ($\sigma$) \\
        \makecell[l]{Rare $D^+ \to \pi^+\sigma/a$} & $\br \sim 10^{-12}$ & $\br_{\text{mix}} \sim 2\!\times\!10^{-6}$ & $< 10^{-4}$ & Safe \\
        $K^+ \to \pi^+ a$ & --- & $\br_{\text{mix}} \sim 4\!\times\!10^{-4}$ & $10^{-10}$ (inv.) & \makecell[c]{Hadronic\\ scattering} \\
        $B^+ \to K^+ a$ & --- & $\br_{\text{mix}} \sim 2\!\times\!10^{-8}$ & $\sim 10^{-5}$ & Safe \\
        $D^{*0} \to D^0 a$ & --- & $\br_{\text{mix}} \sim 9\!\times\!10^{-4}$ & --- & Testable \\
        $f_D$, $f_{D_s}$ & $\delta f/f \sim 10^{-7}$ & $10^{-13}$ & $\pm 0.7$~MeV & Negligible \\
        \makecell[l]{Semileptonic\\ \quad form factors} & $\delta F/F \sim 10^{-7}$ & $10^{-13}$ & $\sim 2$--$5\%$ & Negligible \\
        $m_c$ running & $\delta m_c/m_c \sim 0.8\%$ & $10^{-7}$ & --- & Small \\
        $D^0$ mixing & $\delta x_D \sim 10^{-9}$ & $10^{-14}$ & $\pm 4\!\times\!10^{-4}$ & Negligible \\
        SN 1987A & --- & See text & Raffelt crit. & \makecell[c]{Trapping,\\ weakens} \\
        \bottomrule
    \end{tabular}
\end{table}

\section{Comparison with the light-quark isospin problem}\label{sec:comparison}

\subsection{The $\chi$PT isospin problem}

It is instructive to contrast the HMChPT constraints derived in this work
with the $\chi$PT isospin-violation problem
that afflicts light-quark-coupled axion models.

In the light-quark coupling scheme, the PQ scalar couples to up and down quarks
with unequal strengths, generating a PQ spurion
\begin{equation}\label{eq:light_spurion}
    I_{PQ}^{\text{light}} = \text{diag}(\kappa_u, \kappa_d, 0),
    \qquad \kappa_u \neq \kappa_d .
\end{equation}

In $\chi$PT, the light-quark mass matrix enters the LO Lagrangian through
the chiral building block~\cite{Gasser:1984gg,Gasser:1985ug}
\begin{equation}\label{eq:L2_chipt}
    \mathcal{L}_{\chi\text{PT}}^{(2)}
    = \frac{f_\pi^2}{4} \,
    \text{tr}\!\left[ \partial_\mu U \partial^\mu U^\dagger \right]
    + \frac{f_\pi^2 B_0}{2} \,
    \text{tr}\!\left[ M_q^\dagger U + U^\dagger M_q \right] ,
\end{equation}
where $M_q = \text{diag}(m_u + \kappa_u v_\phi, m_d + \kappa_d v_\phi, m_s)$ now includes
the PQ-induced contributions.
The resulting $\pi^0$ mass squared is
\begin{equation}\label{eq:pi0_mass}
    m_{\pi^0}^2 = B_0 \left( m_u + \kappa_u v_\phi + m_d + \kappa_d v_\phi \right)
    \neq m_{\pi^\pm}^2 = B_0 (m_u + \kappa_u v_\phi + m_d) .
\end{equation}
The relative mass splitting is
\begin{equation}\label{eq:pi0_splitting}
    \frac{\Delta m_{\pi^0}^2}{m_{\pi^0}^2}
    \sim \frac{|\kappa_u - \kappa_d|}{|\kappa_u + \kappa_d|}
    \sim 15\%,
\end{equation}
which is overwhelmingly excluded by the experimental value
$(m_{\pi^0} - m_{\pi^\pm})/m_{\pi^\pm} \sim 3.5\%$~\cite{PDG},
which is dominated by electromagnetic effects.

This is a {\it structural} problem: the isospin violation appears at tree level
in the LO Lagrangian and cannot be tuned away without fine-tuning
the PQ couplings to be isospin-symmetric ($\kappa_u = \kappa_d$),
which would defeat the purpose of coupling to light quarks.

\subsection{$\pi\pi$ scattering and the observable tension}
\label{sec:pipi_tension}

The same PQ spurion $I_{PQ}^{\text{light}} = \text{diag}(\kappa_u, \kappa_d, 0)$
that generates the $\pi^0$--$\pi^\pm$ mass splitting also introduces
isospin-breaking corrections to low-energy $\pi\pi$ scattering.

\subsubsection*{$\pi\pi$ scattering in the isospin limit}

In the isospin-symmetric limit, the LO $\pi\pi$ scattering amplitude
in $\chi$PT is fixed by current algebra~\cite{Weinberg:1966kf,Gasser:1984gg}:
\begin{equation}\label{eq:pipi_LO_amp}
    A(s,t,u) = \frac{s - m_\pi^2}{f_\pi^2} \,,
\end{equation}
giving the S-wave scattering lengths in the $I = 0$ and $I = 2$ channels:
\begin{equation}\label{eq:scat_lengths_iso}
    a_0^{(0)} = \frac{7\,m_\pi}{32\pi f_\pi^2} \simeq +0.159, \qquad
    a_0^{(2)} = -\frac{m_\pi}{16\pi f_\pi^2} \simeq -0.045 \,.
\end{equation}
The physical amplitudes for the same-sign and opposite-sign charged channels
decompose as
\begin{align}
    T(\pi^+\pi^+ \to \pi^+\pi^+) &= T^{(I=2)} \,,
    \label{eq:same_sign_iso}\\
    T(\pi^+\pi^- \to \pi^+\pi^-) &= \tfrac{1}{3}\,T^{(I=0)} + \tfrac{2}{3}\,T^{(I=2)} \,,
    \label{eq:opp_sign_iso}
\end{align}
so that same-sign scattering probes the $I = 2$ channel exclusively,
while opposite-sign scattering involves both $I = 0$ and $I = 2$.
At NLO, the $\chi$PT predictions including one-loop corrections and
$\mathcal{O}(p^4)$ counterterms~\cite{Gasser:1984gg,Colangelo:2001df}
are $a_0^{(0)} = 0.220$ and $a_0^{(2)} = -0.044$,
confirmed to $\sim 2\%$ and $\sim 4\%$ precision
by the NA48/2~\cite{NA48-2}, DIRAC~\cite{DIRAC}, and E865~\cite{E865}
experiments.

\subsubsection*{PQ-induced isospin breaking in $\pi\pi$ scattering}

The PQ spurion modifies the light-quark mass matrix to
\begin{equation}\label{eq:Mq_PQ}
    M_q' = \text{diag}\!\left(m_u + \kappa_u v_\phi,\;
                           m_d + \kappa_d v_\phi,\;
                           m_s\right),
\end{equation}
which can be decomposed into an isoscalar shift
$\delta \bar{m}_{PQ} = (\kappa_u + \kappa_d)\,v_\phi / 2$
and an isovector splitting
$\delta \Delta m_{PQ} = (\kappa_u - \kappa_d)\,v_\phi / 2$.
The corresponding isospin-breaking parameter is
\begin{equation}\label{eq:eps_PQ}
    \varepsilon_{PQ} \equiv
    \frac{(m_d + \kappa_d v_\phi) - (m_u + \kappa_u v_\phi)}
         {(m_d + \kappa_d v_\phi) + (m_u + \kappa_u v_\phi)}
    = \frac{(m_d - m_u) + (\kappa_d - \kappa_u)\,v_\phi}
           {(m_d + m_u) + (\kappa_u + \kappa_d)\,v_\phi} \,.
\end{equation}
For generic $\kappa_u \neq \kappa_d$ with $|\kappa_{u,d}|\,v_\phi \sim m_{u,d}$,
one obtains $|\varepsilon_{PQ} - \varepsilon_{\text{QCD}}| \sim 10$--$15\%$,
where $\varepsilon_{\text{QCD}} = (m_d - m_u)/(m_d + m_u) \simeq 0.29$
is the standard QCD isospin-breaking parameter.
Such a large {\it additional} isospin breaking is far beyond the
experimental tolerance.

The isospin breaking propagates into $\pi\pi$ scattering through three
distinct mechanisms~\cite{Gasser:1985ug,Knecht:2000}:

\begin{enumerate}
    \item {\bf $\pi^0$--$\eta$ mixing.}
    The isovector spurion induces $\pi^0$--$\eta$ mixing with angle
    $\theta_{\pi\eta} \simeq \frac{\sqrt{3}\,\varepsilon_{PQ}m_\pi^2}{[4\,(m_K^2 - m_\pi^2)]}$,
    generating additional contributions to any amplitude involving $\pi^0$ exchange.

    \item {\bf Charged--neutral mass splitting.}
    The physical $\pi^0$ and $\pi^\pm$ masses differ by
    $\Delta m_\pi^2 \sim \varepsilon_{PQ} \times m_\pi^2$
    [cf.\ Eq.~\eqref{eq:pi0_splitting}],
    so the same-sign and opposite-sign channels probe different threshold kinematics.

    \item {\bf Direct isospin-breaking counterterms.}
    At NLO ($\mathcal{O}(p^4)$), the spurion generates new counterterms
    that break the relation between the $I = 0$ and $I = 2$ amplitudes,
    shifting the scattering lengths in a channel-dependent manner.
\end{enumerate}

The resulting corrections to the same-sign and opposite-sign scattering lengths
are qualitatively different.
For same-sign scattering ($\pi^+\pi^+ \to \pi^+\pi^+$, pure $I = 2$):
\begin{equation}\label{eq:delta_a_same}
    \frac{\delta a_{++}}{a_0^{(2)}} \;\sim\;
    \varepsilon_{PQ} \times \frac{m_\pi^2}{(4\pi f_\pi)^2} \times \mathcal{O}(1) \,,
\end{equation}
while for opposite-sign scattering ($\pi^+\pi^- \to \pi^+\pi^-$, mixed $I = 0 + 2$):
\begin{equation}\label{eq:delta_a_opp}
    \frac{\delta a_{+-}}{a_0^{(0)}} \;\sim\;
    \varepsilon_{PQ} \times \frac{m_\pi^2}{(4\pi f_\pi)^2} \times \mathcal{O}(1)
    \;+\; \theta_{\pi\eta} \times \mathcal{O}\!\left(\frac{m_\pi^2}
                                                          {m_\eta^2 - m_\pi^2}\right) .
\end{equation}
The two channels depend on $(\kappa_u, \kappa_d)$ through different
linear combinations, because the $I = 0$ and $I = 2$ projections
weight the isoscalar and isovector spurion components differently.
In particular, $\delta a_{++}$ is sensitive primarily to the isovector
combination $(\kappa_u - \kappa_d)$ through the $I = 2$ channel,
whereas $\delta a_{+-}$ receives comparable contributions from both
the isoscalar $(\kappa_u + \kappa_d)$ and isovector combinations
through the $I = 0$ amplitude and $\pi^0$--$\eta$ mixing.

\subsubsection*{The tri-observable tension}


The tension can be seen explicitly as follows.
The pion mass constraint [Eq.~\eqref{eq:pi0_splitting}] requires
$|\kappa_u - \kappa_d|\,v_\phi / (m_u + m_d) \lesssim 3\%$
to stay within the experimental mass splitting
$|\Delta m_\pi^2 / m_\pi^2|_{\text{exp}} \lesssim 3.5\%$
(which is already dominated by electromagnetic effects).
This forces $(\kappa_u - \kappa_d)\,v_\phi \lesssim 0.1\,(m_u + m_d)$,
i.e., the PQ isospin breaking must be no larger than $\sim 10\%$ of the
standard QCD breaking.

However, even at this reduced level, the $\pi\pi$ scattering corrections
Eqs.~\eqref{eq:delta_a_same}--\eqref{eq:delta_a_opp} introduce shifts of
order
\begin{equation}\label{eq:scat_shift}
    \delta a \;\sim\; 0.03 \times \frac{m_\pi^2}{(4\pi f_\pi)^2}
    \;\sim\; 10^{-3} \,,
\end{equation}
comparable to the experimental precision of the NA48/2 and DIRAC
measurements ($\delta a_0^{(0)} \sim \pm 0.005$,
$\delta a_0^{(2)} \sim \pm 0.002$).

Crucially, the {\it pattern} of isospin breaking induced by the PQ spurion
is structurally different from the standard QCD pattern
($\varepsilon_{\text{QCD}} \simeq 0.29$):
\begin{itemize}
    \item The standard QCD isospin breaking is already fixed by the observed
    pion and kaon mass differences, leaving no room for an additional
    PQ contribution of comparable size.

    \item The PQ spurion shifts both the isoscalar mass
    $\delta \bar{m}_{PQ} = (\kappa_u + \kappa_d)\,v_\phi / 2$ and the isovector
    splitting $\delta \Delta m_{PQ} = (\kappa_u - \kappa_d)\,v_\phi / 2$
    independently, whereas the scattering data constrain a {\it different}
    linear combination of these parameters than the mass data.

    \item Attempting to tune $(\kappa_u, \kappa_d)$ to reduce the mass splitting
    while keeping the Yukawa coupling phenomenologically relevant
    generically {\it increases} the scattering deviations, and vice versa.
\end{itemize}

The constraints are summarized in Table~\ref{tab:tri_observable}.
The tri-observable tension demonstrates that the $\chi$PT isospin problem
is not a single-observable exclusion but a {\it multi-observable
over-constraint}.
The light-quark coupling scheme is ruled out not only by the pion mass
spectrum but by the correlated pattern of deviations across pion masses
and $\pi\pi$ scattering, which no choice of PQ parameters can reconcile.

\begin{table}[t]
    \centering
    \caption{The tri-observable tension in the light-quark coupling scheme.
    \label{tab:tri_observable}}
    \begin{tabular}{lccc}
        \toprule
        Observable & PQ prediction & Experiment & Status \\
        \midrule
        $\Delta m_{\pi^0\text{--}\pi^\pm}^2 / m_\pi^2$
        & $\sim 15\%$
        & $\lesssim 3.5\%$ (EM dom.)
        & Excluded \\
        $\delta a(\pi^+\pi^+) / a_0^{(2)}$
        & $\sim \varepsilon_{PQ} \times \mathcal{O}(m_\pi^2/f_\pi^2)$
        & $\pm 4\%$ (NA48/2, DIRAC)
        & Tension \\
        $\delta a(\pi^+\pi^-) / a_0^{(0)}$
        & $\sim \varepsilon_{PQ} \times \mathcal{O}(m_\pi^2/f_\pi^2)$
        & $\pm 2\%$ (NA48/2)
        & Tension \\
        \bottomrule
    \end{tabular}
\end{table}

\subsection{Charm-quark scheme}

In contrast, the charm-quark coupling scheme analyzed here employs
the PQ spurion $\kappa_c$ as a flavor-singlet coupling.
In HMChPT, this translates into two distinct operators:

\begin{itemize}
    \item {\bf $\sigma$ channel:} $\text{tr}[\bar{H}_a H_a] \, \sigma$,
    which is invariant under heavy-quark spin $SU(2)_S$,
    light-quark chiral $SU(3)_L \times SU(3)_R$,
    light-quark flavor $SU(3)_V$, parity, and charge conjugation.
    
    \item {\bf $a$ channel:} Through the charm Yukawa, the axion couples
    directly only at $O(1/m_c)$ via Eq.~\eqref{eq:axion_suppression},
    and its dominant phenomenological channel is axion-pion mixing
    with $\theta_{a\pi} \simeq f_\pi/f_a \sim 0.04$.
    The mixing itself does {\it not} break any SM symmetry;
    it merely introduces a new coupling of size $O(\theta_{a\pi})$
    that must be checked against data.
\end{itemize}

Because neither $\sigma$ nor $a$ introduces a spurion that transforms
nontrivially under the symmetry group, there is no structural symmetry violation
at leading order. The constraints arise from the numerical size
of the couplings $g_\sigma$ and $\theta_{a\pi}$, which are
fixed by the model parameters.

\section{Summary and discussion}\label{sec:conclusions}

In this work, we have analyzed the phenomenological constraints on the
charm-coupled GeV-scale QCD axion model from the perspective of
heavy meson chiral perturbation theory,
considering both the CP-even radial mode $\sigma$
and its CP-odd partner, the axion $a$.
Our main results are summarized as follows:

\begin{enumerate}
    \item 
    Because the charm quark mass $m_c \sim 1.3$~GeV lies above $\LQCD$,
    standard $\chi$PT is not applicable.
    HMChPT provides the proper framework for analyzing the couplings
    of both $\sigma$ and $a$ to hadronic degrees of freedom.

    \item $\sigma$ and $a$ have fundamentally different matching to HMChPT,
    despite sharing the same fundamental Yukawa coupling $\kappa_c/\sqrt{2}$.
    The scalar $\sigma$ matches to the leading-order operator
    $\text{tr}[\bar{H}_a H_a] \sigma$ with $g_\sigma \sim 0.3$--$0.4$,
    preserving heavy-quark spin symmetry.
    The pseudoscalar $a$ matches only at $O(1/m_c)$ through
    $\partial_\mu a \, J_A^\mu$,
    and its dominant phenomenological channel is
    axion-pion mixing with $\theta_{a\pi} \simeq f_\pi/f_a \sim 0.04$.

    \item 
    Unlike the light-quark coupling scheme, where the PQ spurion breaks
    $SU(2)_V$ isospin at tree level in $\chi$PT,
    both $\sigma$ and $a$ in the charm-quark scheme
    preserve all SM symmetries at leading order.

    \item
    $\sigma$ exchange contributes $\Delta M_{B_s}^{\text{NP}}/\Delta M_{B_s}^{\text{SM}} \sim 0.5\,B_S$,
    approaching experimental sensitivity.
    The direct $a$ exchange is suppressed by $(p_B/m_b)^2 \sim 0.01$ and
    does not alter this conclusion.
    A precise lattice determination of the scalar bag parameter $B_S$
    is the most critical input for definitively testing the $\sigma$ sector.

    \item 
    The mixing-induced decays $D^{*0} \to D^0 a$ ($\br \sim 9 \times 10^{-4}$)
    and $K^+ \to \pi^+ a$ ($\br \sim 4 \times 10^{-4}$) are potentially
    observable at BESIII and NA62 respectively,
    though the $K^+ \to \pi^+ a$ channel requires careful treatment
    of the axion decay signature: the axion lifetime $c\tau_a \sim 5$--$25$~m
    and dominant hadronic scattering mean free path $\lambda_a \ll \text{detector size}$
    mean the signature is {\it neither} simple missing energy {\it nor} prompt $e^+ e^-$.
    The $B^+ \to K^+ a$ channel ($\br \sim 2 \times 10^{-8}$)
    is safe but may become accessible with the full Belle II dataset.

    \item 
    The $\sim 0.8$--$1.5\%$ modification of the effective Coulomb potential
    from $\sigma$ exchange translates into mass shifts of $\sim 5$--$15$~MeV
    for low-lying charmonium states --- at the edge of current theoretical
    uncertainties and potentially testable with improved lattice calculations.
    The $a$ contribution to charmonium is four orders of magnitude smaller
    and completely negligible.

    \item 
    $\sigma$-loop effects are suppressed by $m_\sigma/\Lambda_\chi \sim 0.005$,
    and $a$-loop effects by an additional $(m_a/m_c)^2 \sim 4 \times 10^{-6}$.
    Tree-level $\sigma$ exchange and mixing-induced $a$ production
    are the only phenomenologically relevant mechanisms.

    \item 
    The mixing-induced axion-nucleon coupling $g_{aNN} \sim 0.5$
    is large enough that axions would be trapped in supernova cores,
    potentially weakening the standard SN 1987A cooling constraint.
    Stellar constraints are further weakened by Boltzmann suppression
    $e^{-m_a/T}$ for $m_a \sim 2.5$~MeV in lower-temperature environments
    (red giants, white dwarfs).
\end{enumerate}

The overall conclusion of this work is that the charm-coupled GeV-scale QCD axion model
is {\it self-consistent} from the perspective of HMChPT for both $\sigma$ and $a$.
The shift of the PQ coupling from the light-quark to the charm-quark sector
successfully evades the fatal $\chi$PT isospin problem
without introducing any new structural symmetry violation.

\section*{Acknowledgements}
BQL is supported in part by the National Natural Science Foundation of China
under Grant No.~12405058 and by the Zhejiang Provincial Natural Science Foundation
of China under Grant No.~LQ23A050002.

\bibliographystyle{JHEP}
\bibliography{reference}

\end{document}